\documentclass[11pt,a4paper]{article}

\usepackage[margin=2.5cm]{geometry}
\usepackage{newtxtext}
\usepackage{amsmath,amssymb}
\usepackage{booktabs}
\usepackage{tabularx}
\usepackage{longtable}
\usepackage{multirow}
\usepackage{makecell}
\usepackage{graphicx}
\usepackage{float}
\usepackage[table]{xcolor}
\usepackage{caption}
\usepackage{authblk}
\usepackage{enumitem}
\usepackage{titlesec}
\usepackage[hidelinks]{hyperref}
\usepackage{url}
\usepackage{siunitx}

\setlist{itemsep=0.2em,topsep=0.3em}
\definecolor{accent}{HTML}{2A78D6}
\definecolor{warn}{HTML}{EB6834}
\definecolor{muted}{HTML}{52514E}

\newcommand{\evMeas}{\textcolor{accent}{\textbf{[measured]}}}

\newcommand{\evDoc}{\textcolor{muted}{\textbf{[documented]}}}

\newcommand{\us}{\mbox{\textmu s}}
\newcommand{\code}[1]{\texttt{\detokenize{#1}}}
\DeclareRobustCommand{\fpath}[1]{\path{#1}}

\title{\vspace{-1.2cm}\textbf{Characterizing CUDA Locality Domains on Blackwell GPUs}}

\author[ ]{Rocky Wang\protect\footnotemark[1] \quad Joyce Zhou\protect\footnotemark[1]
  \quad Keith Jiang \quad Jason Zhao\protect\footnotemark[2]}
\affil[ ]{Scitix AI}
\date{}

\begin{document}
\maketitle
\footnotetext[1]{Equal contribution.}
\footnotetext[2]{Corresponding author: \href{mailto:jason@scitix.ai}{jason@scitix.ai}}

\begin{abstract}
\noindent
A Blackwell GPU contains two compute dies, and latest CUDA provides locality domains
(LDs) to control memory placement and restrict computation to a single die.
Despite the introduction of these controls, their performance implications and
underlying costs remain largely unexplored. We study LDs from three
perspectives: memory management and access, computation, and communication.
Across eight B300 GPUs, we use a controlled four-state comparison of stock
execution, memory placement alone, compute binding alone, and both controls
together. We also use mismatched configurations as negative controls and
hardware counters to identify the causes of performance changes.

Our measurements show that the benefit of locality depends on how data is
accessed. A cross-die dependent HBM access costs 193.4\,ns more than the same
access served locally, while a cross-die streaming read reaches about 73\% of
the same-die bandwidth, a reduction of about 27\%. In contrast,
cross-die reads add little latency once data is resident in L2, because the data
can be cached in the accessing die's local L2. These differences explain why
memory placement alone improves remote reads by $2.08\times$ for a 1\,MiB
buffer, while localizing the memory of a 256\,MiB device-to-device copy makes it
about 47\% slower. The results also show that different operations require different
controls: compute binding alone speeds up SM-issued collectives by
$1.495\times$, while resident-data GEMMs require both memory placement and
compute binding to achieve a $1.37\times$ speedup. Neither control provides a
consistent benefit to a bandwidth-bound attention kernel once each configuration
is allowed to select its own tile size.

We also identify a substantial overhead in localized memory allocation that emerges beyond 32\,MiB.
The mapping loop of driver processes localized memory one 2\,MiB page
at a time, with 77--87\% of the cost spent on full-chip barriers and TLB
invalidations. By coalescing physically contiguous mappings and deferring
synchronization, we reduce the time for a 1\,GiB \code{cuMemCreate} operation by
$7.02\times$. Together, our results show that LD performance depends on the
interaction between memory placement, compute placement, and access patterns:
locality can accelerate remote reads and computation on resident data, but can
also reduce streaming performance, and its benefits must be weighed against both
hardware and driver costs.
\end{abstract}

\section{Introduction}
\label{sec:intro}

A Blackwell GPU package contains two compute dies, each with its own L2 cache
slices and memory controllers~\cite{nvidia2025localitydomain,liu2026dgna}.
The two dies communicate through an on-package fabric, so an access that
crosses the die boundary can cost more than one that stays within a die.
CUDA 13.4 exposes this hardware topology through locality domains (LDs), giving
programmers more control over where data resides and where computation runs.

LDs provide two independent controls. Memory can be placed on a selected die
instead of being striped across both dies, using virtual memory management (VMM)
or an LD-tagged memory pool~\cite{nvidia2025localitydomain}. Computation can be
restricted to the 70 SMs on a selected die by partitioning SM resources and
creating a green context on that
partition~\cite{nvidia_green_context,bakita2023hardware}. These controls give
programmers a way to make data and computation local to each other. However,
the performance effects of these controls have received little systematic
study, and it is not clear when using them is worthwhile.

Existing guidance on LDs generally points to performance benefits from improved
locality~\cite{nvidia2025localitydomain,nvidia_mlopart}. Our measurements,
however, show that the results depend strongly on how data is accessed.
Placing a 1\,MiB buffer read by a remote GPU increases transfer bandwidth from
75.7 to 157.2\,GB/s, more than doubling throughput. The
same memory placement produces no measurable improvement for a 1\,MiB buffer
used by a local collective. For a 256\,MiB device-to-device copy, it makes
performance 47\% worse. Thus, placing data on one die does not necessarily
make an application faster. To understand why these results differ, we need to
look at which component accesses the data, where that component runs, and how
often data must cross the die boundary.

We study LDs on eight B300 GPUs, covering memory management and access,
computation, and communication. Across these experiments, we use the same four
configurations: stock execution, memory placement alone, compute binding alone,
and both controls together. This comparison lets us separate the effects of the
two controls and determine whether they work independently or need to be
combined. We also use mismatched placement as a negative control, hardware
counters to examine memory behavior, and SM-count controls to distinguish
locality benefits from the effects of using fewer SMs.

To interpret the measurements, we develop a cost model that separates the cost
of a cross-die access into two parts: the latency of each cross-die hop and the
bandwidth cost of moving data. The first matters when an operation depends on
each access completing before the next can proceed; the second matters when
large amounts of data must be transferred. The benefit of LD therefore depends
on both the access pattern and which placement control changes the path of an
access. For example, placing a buffer on the appropriate die can help a remote
GPU read it without changing where the copy engine runs. A local collective can
instead benefit from binding its computation to one die, even when changing
memory placement has little effect. A kernel that repeatedly reads resident
data, such as GEMM weights, may need both controls to work together. The model
turns these observations into testable predictions, which we evaluate using
dependent-load and streaming microbenchmarks, hardware-counter analysis, and
controlled workload experiments.

Our measurements reveal several important differences in LD behavior. Cross-die
dependent HBM accesses cost an additional 193.4\,ns, and cross-die streaming reaches about
73\% of the same-die bandwidth, a reduction of about 27\%. Yet cross-die reads
add little latency once the data is resident in L2: the accessing die can
retain a local copy, while writes continue to cross the fabric on each access.
This difference helps
explain why memory placement benefits some access patterns but not others. We
also find that the two placement controls can complement each other. For
resident-data GEMMs, memory placement alone achieves only $0.75\times$ the
baseline performance, and compute binding alone achieves $1.04\times$, while
using both achieves $1.37\times$. For SM-issued collectives, compute binding
alone delivers a $1.495\times$ speedup, whereas a configuration using a
comparable number of SMs across both dies achieves only $0.999\times$. In
contrast, an apparent $1.27\times$ improvement for a bandwidth-bound attention
kernel disappears when each configuration is allowed to select its own tile
size. These results show why LD performance must be evaluated with controlled
comparisons rather than inferred from locality alone.

We also examine the cost of allocating localized memory. Beyond a 32\,MiB
threshold, allocation overhead becomes substantial. We trace this cost to the
open-source GPU driver's mapping loop, which processes localized memory one
2\,MiB page at a time. Full-chip barriers and TLB invalidations account for
77--87\% of the cost of each call, rather than the page-table updates
themselves. We modify the driver to coalesce physically contiguous mappings and
defer synchronization. This reduces the time for a 1\,GiB \code{cuMemCreate}
operation by $7.02\times$, although some overhead remains. This result shows
that the cost of using LDs comes not only from data access and placement, but
also from the memory-management path used to establish that placement.

Our contributions are as follows:

\begin{enumerate}[leftmargin=1.8em,itemsep=0.25em]

\item \textbf{A cost model for cross-die accesses.}
We develop a two-component model that separates per-hop latency from
data-transfer bandwidth cost. The model predicts when memory placement, compute
binding, or both should improve performance based on the accessor and the
location of the data.

\item \textbf{A controlled measurement methodology.}
We evaluate LDs across memory management and access, computation, and
communication on eight B300 GPUs. By comparing stock execution, memory-only
placement, compute-only binding, and full placement, and by using negative
controls, hardware counters, and SM-count controls, we isolate the effects of
locality from other factors.

\item \textbf{A characterization of LD performance.}
We show that LDs can improve remote-GPU reads and SM-issued collectives, that
resident-data GEMMs benefit from combining memory and compute placement, and
that localized placement can substantially hurt streaming performance. We also
identify read--write asymmetry, the limited effect of cross-die reads on
L2-resident data, and the importance of controlling for kernel configuration
when measuring speedups.

\item \textbf{A driver fix for localized allocation overhead.}
We identify an allocation overhead that emerges beyond 32\,MiB and trace it to
repeated mapping calls and synchronization in the open-source driver. By
coalescing contiguous mappings and deferring synchronization, we reduce the time
for a 1\,GiB \code{cuMemCreate} operation by $7.02\times$.

\end{enumerate}

\section{Background}
\label{sec:background}

\subsection{Blackwell Architecture and Locality-Domain Controls}

A Blackwell GPU contains two compute dies connected by an on-package fabric.
Each die has its own L2 cache slices and memory controllers~\cite{liu2026dgna}.
Accessing data on the other die therefore requires communication across this
fabric. Each B300 GPU exposes two locality domains (LDs), each with 70 SMs out
of 148 SMs in total~\cite{nvidia2025localitydomain}.
The GPU has 126\,MiB of L2 cache in total, so each die exposes about half of it.

SM indices do not correspond to the two locality domains in a simple, contiguous
split. For example, the domain-0 green context includes SMs
\{0--3, 6--9, 16--19, \ldots, 142--147\}, while domain-1 includes
\{4--5, 10--15, \ldots, 130--135\}. A range of SM indices therefore does not, by
itself, tell us which die an SM belongs to. Throughout this paper, we determine
compute placement from the green-context resource configuration and verify where
kernels execute by sampling \texttt{\%smid}.

CUDA provides two ways to control locality. The first is memory placement. With
virtual memory management (VMM), an application can pass an LD location to
\code{cuMemCreate}. It can also use a memory pool tagged with an LD. Both
mechanisms use the location type
\path{CU_MEM_LOCATION_TYPE_DEVICE_LOCALITY_DOMAIN} to specify the target
domain~\cite{nvidia2025localitydomain}. We check the resulting placement by
reading back \code{LOCALITY_DOMAIN_ORDINAL}. On all eight GPUs in our
experiments, this
attribute reports the requested domain. By contrast, \code{cudaMalloc} does not
provide an explicit way to select an LD, and its \code{LOCALITY_DOMAIN_ORDINAL}
attribute reads back as $-1$. Our measurements show that memory allocated this
way is striped across both dies.

The second control is compute placement. CUDA's \code{cuDevSmResourceSplit} API
can split SM resources into groups that correspond to the two locality
domains~\cite{bakita2023hardware}. These groups can then be used to create green
contexts and their associated streams, restricting kernel execution to the
selected SM resources~\cite{nvidia_green_context}. Memory
placement determines where data resides, while compute placement determines
which SMs execute the work. Since these controls are independent, we vary them
separately and together throughout our experiments.

\subsection{Locality Domains and GPU Partitioning Mechanisms}

LDs are not the only way to control how GPU resources are used. NVIDIA also
provides mechanisms such as MIG, MPS, green contexts, and MLOPart. Although some
of these mechanisms can expose or use per-die resource configurations, they
serve different purposes and do not provide the same placement controls.

\begin{itemize}[leftmargin=1.4em,itemsep=0.15em]
    \item \textbf{Multi-Instance GPU (MIG).} MIG divides a GPU into separate
    instances with dedicated resources. A two-instance configuration separates a GPU into
    two instances, each with 70 SMs and 89\,GB of memory,
    matching the SM count of the two locality domains~\cite{nvidia_mig}.
    The key difference is that MIG creates separate GPU instances, whereas LDs
    let an application
    control memory and compute placement without requiring a separate instance
    for each domain.

    \item \textbf{Multi-Process Service (MPS).} MPS allows multiple processes to
    share a GPU and provides controls over the resources available to each
    client~\cite{nvidia_mps}. These controls determine how GPU resources are
    allocated among clients, but resource allocation alone does not specify
    which locality
    domain executes a client's work.

    \item \textbf{Green contexts.} Green contexts let applications create
    execution contexts backed by selected SM resources~\cite{nvidia_green_context}.
    They provide the mechanism we use to restrict computation to one locality
    domain, but they do not determine where memory is allocated.

    \item \textbf{MLOPart.} MLOPart exposes per-die partitions as devices that
    applications can target~\cite{nvidia_mlopart}. Unlike LD placement, which
    controls memory and compute placement within a GPU, MLOPart uses
    partitioned devices as the unit of resource allocation. We separately
    evaluate the runtime and driver overheads associated with concurrent
    partitions in this paper.
\end{itemize}

The main difference between these mechanisms is what they control. MIG and
MLOPart expose partitioned GPU resources, while MPS manages resource sharing
among processes. Green contexts restrict execution to selected SM resources, and
LDs provide separate controls for memory placement and compute placement.
Recent work further shows that die-level structure affects GPU scheduling, and
that thread block clusters let SMs cooperate within a
kernel~\cite{fan2026dissecting,nvidia2025threadblockcluster,luo2026clusterfusion,he2026taming}.
These approaches change which SMs cooperate, but they do not by themselves
control where data resides, which is the placement axis we study. These
differences matter when comparing LDs with other approaches: a performance
change may come from improved locality, but it may also come from changes in
resource capacity or isolation. We return to this distinction in
Section~\ref{sec:discussion}, where we use our cross-die cost model to interpret
the experimental results.

\section{A Cost Model for Cross-Die Access}
\label{sec:model}

\subsection{Cost Model}

To understand when locality domains improve performance, we separate execution
time into the cost of individual accesses and the cost of moving bulk data.
Consider a workload that transfers $S$ bytes and performs $n$ dependent accesses
to data on the remote die. We use the following first-order model:

\begin{equation}
T \;=\; t_0 \;+\; n\,\bigl(t_{\mathrm{hop}} + \delta\bigr) \;+\; \frac{S}{B_{\mathrm{eff}}},
\label{eq:cost}
\end{equation}

Here, $t_0$ represents overheads that do not depend on placement. The term
$t_{\mathrm{hop}}$ is the additional latency of accessing data across dies
rather than locally, and $\delta$ captures the additional cost or savings
associated with how the access is served after the initial remote access. The
effective bandwidth, $B_{\mathrm{eff}}$, determines how long it takes to
transfer the remaining $S$ bytes. This model does not capture every detail of
GPU execution, such as overlapping accesses or hiding latency, but it lets us
examine two distinct sources of cost: the number of accesses and the amount of
data transferred.

We can use the model to compare the default placement, which distributes memory
across both dies, with a placement that localizes memory to one die. Let
$T_{\mathrm{plain}}$ denote the execution time under the default placement and
$T_{\mathrm{local}}$ the time under localized placement. The time saved by
localization is

\begin{equation}
\begin{aligned}
\Delta
&= T_{\mathrm{plain}} - T_{\mathrm{local}} \\
&\approx n\,\Delta t_{\mathrm{access}}
 + S\left(
       \frac{1}{B_{\mathrm{plain}}}
       - \frac{1}{B_{\mathrm{local}}}
     \right),
\end{aligned}
\label{eq:delta}
\end{equation}

where $\Delta t_{\mathrm{access}}$ is the per-access cost removed by
localization, and $B_{\mathrm{plain}}$ and $B_{\mathrm{local}}$ are the
effective bandwidths under the two placements. A positive $\Delta$ means
localization reduces execution time; a negative value means it makes the
workload slower.

The two terms can favor different placements. The first term grows with the
number of accesses whose cost is reduced. The second depends on the amount of
data transferred and the bandwidth available under each placement. If
localization reduces access latency without sacrificing bandwidth, both terms
favor localization. But if placing data on one die reduces the available
bandwidth, the second term can become negative and outweigh the latency savings.
This is why localization may help a workload with frequent dependent accesses
but hurt a kernel that streams a large amount of data.

The model gives us a way to make predictions before examining each workload. Its
two key platform-dependent quantities are the effective per-access cost
associated with crossing dies and the bandwidth under localized placement
relative to the default placement. We measure both experimentally in
Section~\ref{sec:mem}, and use hardware counters to investigate why the costs
differ across access patterns.

\subsection{Placement Hypotheses}

The model estimates the benefit of removing cross-die access costs, but it does
not tell us which LD control can remove them. Memory placement changes where
data resides; compute placement changes where the kernel executes. Which control
matters therefore depends on who issues the access and where the data is
located.

We use this distinction to formulate four hypotheses:

\begin{itemize}[leftmargin=1.4em,itemsep=0.25em]
    \item \textbf{Remote accessors.} When a remote GPU or copy engine reads a
    buffer on the local GPU, placing the buffer on the appropriate die may
    reduce cross-die traffic. Binding local SMs to a die should have little
    effect if those SMs do not issue the relevant accesses. We therefore expect
    memory placement to provide the main benefit.

    \item \textbf{Local SM-based collectives.} When SMs issue accesses as part
    of a collective operation, binding execution to one die may reduce cross-die
    traffic. Changing memory placement alone may not help if the main cost comes
    from where the SMs execute and which cache or memory resources they access.
    We therefore expect compute placement to matter more than memory placement.

    \item \textbf{Repeated accesses to resident data.} A kernel such as GEMM may
    repeatedly read weights that remain resident in memory or cache throughout
    execution. In this case, performance depends on both where the weights
    reside and where the SMs access them from. We expect memory placement and
    compute placement to work together when neither control alone fully aligns
    the data with its accessor.

    \item \textbf{Bandwidth-dominated streaming.} A local kernel that streams a
    large working set may be limited mainly by memory bandwidth rather than
    access latency. If localization reduces the bandwidth available to the
    kernel, that loss may outweigh the benefit of avoiding cross-die accesses.
    We therefore expect the default distributed layout to perform better in some
    bandwidth-bound workloads.
\end{itemize}


\begin{table}[t]
\centering
\footnotesize
\setlength{\tabcolsep}{4pt}
\renewcommand{\arraystretch}{1.05}
\caption{Predicted placement effects and contrary evidence across accessor classes.}
\label{tab:pred}
\begin{tabularx}{\textwidth}{
    @{}
    >{\raggedright\arraybackslash}p{2.6cm}
    >{\raggedright\arraybackslash}p{4.0cm}
    >{\raggedright\arraybackslash}X
    @{}
}
\toprule
\textbf{Accessor class} &
\textbf{Prediction} &
\textbf{Contrary evidence} \\
\midrule

Remote reader &
Memory placement helps; compute placement alone is neutral. &
Compute placement alone provides a comparable gain, or memory placement
does not help. \\

Local SM collective or copy &
Compute placement helps; memory placement alone does not. &
Memory placement alone provides a comparable gain, or the gain is explained
by SM count rather than die affinity. \\

Local SM reading resident data &
Combined placement outperforms either control alone. &
Either control alone achieves a comparable gain, or the combination adds
no benefit. \\

Bandwidth-bound streaming &
Localization hurts at large working-set sizes. &
Localization continues to help at large sizes without a corresponding
bandwidth explanation. \\

\bottomrule
\end{tabularx}
\end{table}

These hypotheses are predictions, not assumptions that localization will help.
Each predicts which control should matter for a particular access pattern, and
can be rejected if the measurements show otherwise. Table~\ref{tab:pred}
summarizes the predictions and the observations that would contradict them.

\subsection{Alternative Explanations and Controls}

A measured speedup does not necessarily mean that locality is responsible.
Changing an LD configuration can also change the number of SMs available to a
kernel, its launch configuration, or the effective memory bandwidth. We design
the experiments in Section~\ref{sec:protocol} to separate these effects.

\textbf{Resource allocation rather than placement.}
An LD configuration may use only 70 SMs, compared with 148 SMs on the full GPU.
A speedup or slowdown could therefore result from changing the number of SMs
rather than from keeping computation on one die. To test this possibility, we
compare configurations with approximately the same SM count, placing the
selected SMs either on one die or across both dies.

\textbf{Kernel configuration rather than placement.}
Different placement configurations may use different generated code or launch
settings, which can affect performance independently of locality. We therefore
compare configurations using the same binary where possible and check their
launch configurations. For workloads where tile selection affects performance,
we also compare configurations after allowing each to choose its own tile size.

\textbf{Bandwidth effects rather than per-access costs.}
A speedup at one working-set size does not, by itself, show that localization
has reduced access latency. The result could instead reflect a change in
effective bandwidth. We therefore sweep working-set sizes and compare the
results with the model's two terms. If the benefit changes as the amount of data
transferred grows, this helps determine whether bandwidth loss offsets the
savings from avoiding cross-die accesses.

These controls help us determine whether the observed performance changes are
consistent with the model or better explained by other factors.
Section~\ref{sec:res:smcount} presents the SM-count control results, while the
remaining experiments examine the effects of access pattern, working-set size,
and placement choice.

\section{Measurement Design}
\label{sec:protocol}

\subsection{Platform and Software Stack}

Unless stated otherwise, all measurements run on the same platform: eight B300
SXM6 AC GPUs (compute
capability 10.3), with driver 615.71.09, CUDA 13.4, NCCL 2.32.3, and nccl-tests
2.20.0. We use Nsight Compute 2026.3.1 for profiling. The one exception is the
static-BAR1 fix evaluation in Section~\ref{sec:mem:rootcause}, which was
collected on a B200; we flag that comparison where it appears. MIG cannot be
enabled
inside our container, so we do not report direct measurements of MIG
performance. Instead, we use NVIDIA's documentation and a substitute experiment
to examine the relevant resource-partitioning question~\cite{nvidia_mig}, and
distinguish these results from direct MIG measurements.

The experiments test the four placement hypotheses from Section~\ref{sec:model}.
We first compare the effects of memory placement and compute placement, then use
targeted workloads and additional controls to check whether the results are
consistent with the model or can be explained by other factors.

\subsection{The Four States}

We compare four placement states and one negative control, summarized in
Table~\ref{tab:states}. Each comparison changes the placement controls while
keeping the other experimental conditions fixed as far as possible.

\begin{table}[H]
\centering\small
\caption{The four-state switch model used throughout. Any pair of numbers in this report differs
only in the state applied, never in the binary or the protocol.}
\label{tab:states}
\begin{tabularx}{0.86\textwidth}{@{}llX@{}}
\toprule
\textbf{State} & \textbf{Memory} & \textbf{Compute} \\
\midrule
Stock        & \code{cudaMalloc} (striped across both dies) & all 148 SMs \\
Memory-only  & VMM or pool allocation placed in a chosen domain & all 148 SMs \\
Compute-only & striped across both dies & green context pinned to one domain (70 SMs) \\
Full-LD      & chosen domain & pinned to the same domain \\
Mismatch     & domain $X$ & pinned to domain $Y \neq X$ \\
\bottomrule
\end{tabularx}
\end{table}

Two comparisons are particularly important. First, the \emph{compute-only}
configuration uses 70 SMs, matching the SM count of the \emph{full-placement}
configuration; the \emph{memory-only} configuration keeps all 148 SMs, as
Table~\ref{tab:states} shows. Comparing compute-only with full placement
therefore isolates what memory placement adds once compute is already confined
to one die, at a fixed SM count. This comparison alone does not rule out
every effect of resource distribution, so we also run the SM-count control
described below.

Second, the \emph{mismatch} configuration places data on one die but binds
computation to the other. It keeps the kernel, memory footprint, and SM count
unchanged while deliberately separating data placement from compute placement.
If a performance benefit depends on bringing data and its accessor together,
this configuration should behave differently from one in which they are placed
consistently.

\subsection{Workloads Chosen to Isolate Accessor Classes}

We use one representative workload for each accessor class, together with
microbenchmarks that measure the underlying access costs
(Table~\ref{tab:workloads}). The microbenchmarks measure the two quantities in
Equation~\eqref{eq:delta} separately. A dependent pointer chase measures access
latency with little memory-level parallelism, allowing us to compare local and
cross-die accesses. Separate streaming kernels measure effective bandwidth and
distinguish read and write behavior, which a combined copy operation could hide.

\begin{table}[H]
\centering\footnotesize
\caption{Experimental workloads, why each was chosen, and the dimensions swept.}
\label{tab:workloads}
\begin{tabularx}{\textwidth}{@{}l>{\raggedright\arraybackslash}p{6.3cm}>{\raggedright\arraybackslash}X@{}}
\toprule
\textbf{Layer} & \textbf{Workload and rationale} & \textbf{Dimensions} \\
\midrule
Primitive & Self-written pointer chasing plus separate streaming read/write kernels; the only
measurement that directly separates accessor die from data die & accessor $\{d0,d1,\text{unbound}\}$
$\times$ data $\{d0,d1,\text{interleaved}\}$ $\times$ 64\,KB--256\,MB $\times$ read/write \\
Management & Per-API timing of seven VMM calls, plus a memory pool & 1\,KiB--1\,GiB, 30--66\,MiB in
2\,MiB steps, cold and warm pools, 1--8 GPUs $\times$ 1--8 threads \\
Copy & nvbandwidth, which exercises copy engines and SMs as distinct accessors & local D2D and
remote $\times$ CE/SM $\times$ 64\,KiB--256\,MiB $\times$ four states \\
Communication & nccl-tests (general library) and flash-ar/flash-ag (a library that can delete
algorithm phases) & AR/AG/A2A $\times$ 2/4/8 ranks $\times$ R0/R2 $\times$ 1\,KB--64\,MB $\times$
four states \\
Compute & DeepGEMM FP8/FP4 masked and normal grouped GEMM, plus decode-attention kernels used as
the bandwidth-bound counterexample & M=1--16, hot/cold $\times$ four states plus mismatch; decode
sweeps $S$, $B$ and tile size \\
Comparison & MPS at 25/50/75/100\%, MLOPart, MIG (documentation only) & \texttt{\%smid} landing, side-by-side
comparison, skewed load \\
\bottomrule
\end{tabularx}
\end{table}

Each application workload tests a different placement hypothesis.

For remote accessors, we use the copy-engine (CE) cases in nvbandwidth. In these
tests, the copy engine initiates the transfer rather than an SM. They let us
examine whether changing the location of the source buffer helps a remote GPU
read it, without relying on compute placement to change the access path.

For SM-issued communication, we use nccl-tests in symmetric-memory registration
mode. These tests exercise the relevant collective communication path at
different rank counts. We inspect the launch configurations and compare
placement states to determine whether binding computation to one die improves
performance.

For repeated accesses to resident data, we use DeepGEMM's grouped GEMM. Its
eight MoE groups can be divided evenly across the two dies, allowing us to
compare memory placement and compute placement separately and together when the
kernel repeatedly accesses its weights.

We also test a case where the model predicts little benefit. Decode attention is
bandwidth-bound, and the split configuration studied here has only one cross-die
stage: the LSE reduction. Once the kernel approaches the bandwidth limit,
removing this cross-die stage may have little effect on total execution time.
Section~\ref{sec:res:attn} tests this prediction. We also allow each
configuration to select its own tile size. The apparent $1.27\times$ gain
disappears under this comparison, showing that the initial difference came from
tile selection rather than a consistent benefit from locality.

\subsection{Controls for Competing Explanations}

\textbf{Resource count.}
The configurations that bind compute to a die use 70 SMs, but matching the SM
count does not by itself
show that a speedup comes from die affinity. We therefore use MPS at 50\% and
25\% resource allocations to test configurations with a comparable number of SMs
distributed across both dies.

Before running the experiment, we set the decision rule in the analysis script.
If the MPS control recovers at least the predefined $1.25\times$ speedup
threshold, the result favors the explanation that resource allocation accounts
for the gain. If performance remains near $1.0\times$, the result supports the
explanation that keeping computation on one die matters. This comparison helps
distinguish the two explanations, although it does not eliminate every possible
difference in resource distribution.

\textbf{Kernel and launch configuration.}
A placement comparison can be misleading if the configurations also use
different kernels or launch settings. For comparisons using the same binary, we
record the grid and block dimensions in each state and check whether they
change. Matching launch configurations rules out differences in grid and block
geometry as the direct explanation for a speedup, although it does not rule out
every difference in execution behavior or resource allocation.

\textbf{Bandwidth effects.}
We vary message size or batch size in the application experiments to see how the
placement benefit changes with the amount of data transferred. The model
predicts that the latency-related benefit depends on the number of affected
accesses, whereas the bandwidth cost grows with the volume of data. If
increasing the transfer size eventually reduces or reverses the benefit, and the
crossover agrees with the measured latency and bandwidth terms, this supports
the model. A trend that these measurements cannot explain would count against
it.

\subsection{Threats to Validity}

Our experiments have two main limitations.

\emph{Limited access to collective-path counters.}
Both Nsight Compute replay modes fail when used with NCCL's memory management in
our container. We therefore cannot directly use fabric-traffic counters to
establish why the collective-path performance changes. Instead, we rely on the
SM-count control and launch-configuration comparisons. We treat these results as
indirect evidence and distinguish them from mechanisms that we confirm directly
with hardware counters.

\emph{MIG cannot be instantiated.}
We cannot enable MIG in our experimental environment. Our comparison with MIG
therefore relies on NVIDIA's documentation~\cite{nvidia_mig} and a substitute
experiment that tests the relevant resource-partitioning premise. We do not
present the substitute experiment as a direct measurement of MIG performance.


We also take several steps to keep measurements reproducible. A global lock
serializes experiments because concurrent bandwidth probes interfere with one
another. We bind each GPU to its corresponding NUMA node, using node
$\lfloor i/2 \rfloor$ for GPU $i$. For every LD allocation, we read back the
locality-domain ordinal and discard the allocation if the reported domain does
not match the requested one.

\section{Memory Placement and Access}
\label{sec:mem}

\subsection{Research Questions}

Equation~\eqref{eq:delta} describes two ways in which memory placement can
affect execution time. It can reduce the cost of individual accesses, but it can
also change the effective bandwidth available to a workload. Which effect
matters more depends on the access pattern. This section measures both costs,
uses hardware counters to examine the difference between reads and writes, and
tests whether memory placement helps remote accessors as predicted by our model.

We also measure the cost of allocating localized memory. Unlike the cost of
accessing data, allocation overhead is paid when memory is created or released.
We measure this overhead separately and determine how much reuse is needed to
recover it.

\subsection{Method}
\label{sec:mem:method}

An end-to-end copy time does not tell us how much of the cost comes from access
latency and how much comes from bandwidth. It also combines the behavior of the
accessor with that of the memory being accessed. We therefore use two
microbenchmarks to measure these costs separately.

\textbf{Dependent-load latency.}
We use a pointer chase in which each load depends on the result of the previous
load. This dependency prevents the kernel from issuing many independent memory
requests at once, making the result more sensitive to access latency than to
aggregate throughput~\cite{liu2026dgna}. The kernel uses cache-global loads that
bypass L1, and we sweep working-set sizes from 64\,KB to 256\,MB.

A green context binds the accessor to a selected die, while VMM places the data
on a selected die. We also use interleaved allocations from \code{cudaMalloc} as
a baseline. By testing all combinations of accessor and data placement,
including working sets both smaller and larger than one die's L2 capacity, we
can distinguish accesses served from cache from those that reach remote DRAM.

\textbf{Streaming read and write.}
We use separate kernels to stream data in each direction. This avoids combining
reads and writes in a single \code{memcpy}, where one direction's behavior could
obscure the other. Each kernel processes a 256\,MB buffer, well beyond L2
capacity, to measure the steady-state effective bandwidth $B_{\mathrm{eff}}$ in
Equation~\eqref{eq:cost}.

We report the median across three independent process runs. For every LD
allocation, we read back the locality-domain ordinal and discard the allocation
if it does not match the requested domain.

\subsection{Results}
\label{sec:mem:access}

\begin{figure}[H]
  \centering
  \includegraphics[width=0.68\textwidth]{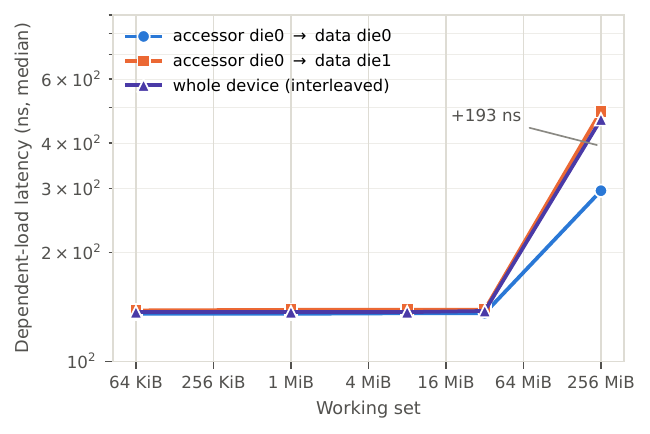}
  \caption{Dependent-load latency (\code{load.global.cg}, bypassing L1; median of three runs).
  The three curves coincide while the working set fits in one die's L2; only at 256\,MB, where
  every access must come from HBM, does the cross-die curve separate.}
  \label{fig:latency}
\end{figure}

\begin{table}[H]
\centering\small
\caption{Primitive-level key points, accessor pinned to die~0; ``interleaved'' is the stock
\code{cudaMalloc} layout. Latency in ns, bandwidth in GB/s; each cell is a median of three
independent processes.}
\label{tab:latbw}
\begin{tabular}{@{}llrrr@{}}
\toprule
\textbf{Metric} & \textbf{Working set} & \textbf{Same die} & \textbf{Cross die} & \textbf{Interleaved} \\
\midrule
\multirow{2}{*}{Read latency} & 8\,MB (fits L2) & 135.8 & 138.8 & 137.3 \\
                              & 256\,MB (HBM) & 295.3 & \textbf{488.8} & 466.1 \\
\midrule
\multirow{2}{*}{Read bandwidth} & 8\,MB & 7592.6 & 7549.7 & 7612.2 \\
                                & 256\,MB & 3628.1 & \textbf{2645.7} & \textbf{5229.8} \\
\midrule
\multirow{2}{*}{Write bandwidth} & 8\,MB & 3699.4 & \textbf{2439.7} & 3678.8 \\
                                 & 256\,MB & 3219.5 & 2351.6 & 3795.2 \\
\bottomrule
\end{tabular}
\end{table}

Table~\ref{tab:latbw} and Figure~\ref{fig:latency} show three main results.

\textbf{Cross-die latency increases sharply when accesses reach remote DRAM.}

For the 256\,MB working set, dependent-load latency is 295.3\,ns for same-die
access and 488.8\,ns for cross-die access. With the accessor bound to die~0 the
increment is 193.4\,ns; the reverse direction gives 191.5\,ns, and the
average over both directions is 192.5\,ns. Unless stated otherwise, we quote
the die-0 direction. At 8\,MB, the difference is only 3.0\,ns, and it remains
close to that value at every L2-resident size we measured: 2.8--3.0\,ns when
averaged over both accessor directions, or 2.7--3.3\,ns across the eight GPUs
at 8\,MB (Section~\ref{sec:res:symmetry}).

The result is similar when we reverse the accessor and data dies: the additional
latency is 191.5\,ns. Averaged over the two accessor directions per GPU, the
measured increment ranges from 191.4 to 193.3\,ns across all eight GPUs. The
per-direction values span a slightly wider band (190.9--194.9\,ns). The large
penalty therefore appears when dependent accesses reach remote DRAM; crossing a
die boundary does not add the same cost to every access.

\textbf{Reads and writes respond differently to cross-die placement.}

For L2-resident data, cross-die reads add little measured latency. Writes behave
differently: cross-die writes reduce bandwidth from 3699.4 to 2439.7\,GB/s, a
34.1\% decrease. A single direction-independent cost cannot explain both
results. We use hardware counters below to examine where the difference comes
from.

\begin{figure}[H]
  \centering
  \includegraphics[width=0.95\textwidth]{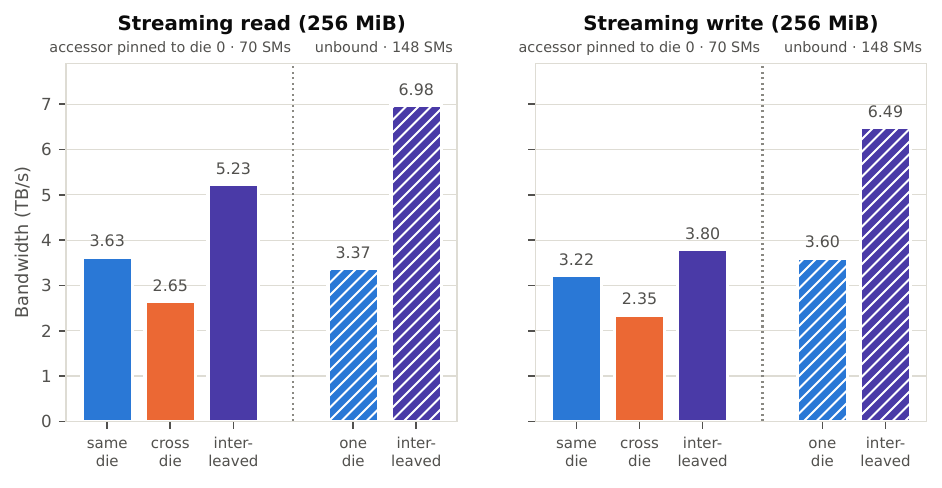}
  \caption{Steady-state bandwidth at 256\,MiB. Bars left of the divider use an accessor pinned
  to die~0, matching Table~\ref{tab:latbw}; there, cross-die read and write both fall to about
  73\% of the same-die figure, while the interleaved layout lets the pinned accessor draw on
  both dies' channels. Hatched bars right of the divider use an unbound accessor on all 148 SMs
  and are shown only as reference ceilings; they mix in a change in SM count, so they must not be
  compared against the pinned bars as if they differed in layout alone.}
  \label{fig:bw}
\end{figure}

\textbf{Restricting memory to one die can reduce streaming bandwidth.}

For the 256\,MB streaming workload, the interleaved layout reaches 5229.8\,GB/s,
compared with 3628.1\,GB/s when memory is localized to one die. Interleaving
allows the workload to use memory channels on both dies, while localization
restricts the available bandwidth to one die. The bandwidth loss can outweigh
the savings from avoiding cross-die accesses, particularly when the workload
transfers large amounts of data.

The comparison also depends on which SMs issue the accesses. The 5229.8\,GB/s
figure uses an accessor bound to die~0, matching the placement states in
Table~\ref{tab:latbw}. An unbound accessor running on all 148 SMs reaches
6975.1\,GB/s on the same interleaved buffer, close to the 7271.1\,GB/s obtained
by summing the two dies' separately measured local-accessor bandwidths (3628.1
and 3643.0\,GB/s); running both dies' local accessors concurrently measures
7210.4\,GB/s. The gap between layouts persists at
full SM count: with the same unbound accessor, confining the buffer to one die
drops reads to 3373.4\,GB/s, roughly half of 6975.1. We report the
bound-accessor values in the table because they isolate placement from SM count,
and Figure~\ref{fig:bw} shows both accessor conditions side by side so the two
effects are not conflated. The unbound interleaved figure is the whole-device
ceiling used for the MPS comparison in Section~\ref{sec:discussion}.

\subsection{Read and Write Traffic Across Dies}
\label{sec:mem:counter}

The hardware counters help explain why reads and writes behave differently. The
B300 does not expose a counter explicitly named for locality domains or dies. We
therefore use \path{lts__t_sectors_srcunit_ltcfabric_op_read.sum} and its write
counterpart, which count sectors received from the LTC fabric. Across the four
accessor--data placement combinations, these counters show different patterns
for reads and writes.

\begin{figure}[H]
  \centering
  \includegraphics[width=0.6\textwidth]{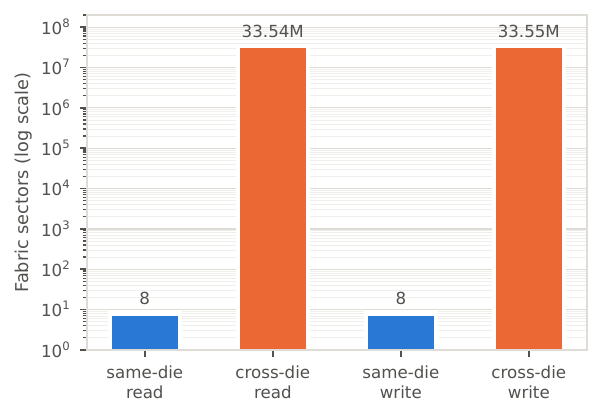}
  \caption{Cross-die fabric sectors for a single 256\,MiB launch, on a log axis. Same-die traffic
is not zero: it is 8 sectors against 33.5M for cross-die traffic, seven orders of magnitude
apart, which is why the same-die bars are invisible on a linear scale. All four same-die
combinations read exactly 8 and write exactly 0, so the residue is a fixed per-launch overhead
rather than data movement. Cross-die traffic equals the full byte count
(33.54M sectors $\times$ 32\,B $=$ 1.073\,GB $\approx$ 1\,GiB, i.e.\ one fabric crossing per byte
transferred).}
\label{fig:fabric}
\end{figure}

\textbf{Cross-die reads can be served from the accessor's local L2 after the initial access.}
For an 8\,MiB cross-die read, we observe only 13,625 L2 misses and eight
\path{op_read} fabric sectors, out of 33.56M L2 lookups that hit. The aggregate
\path{lts__t_sectors_srcunit_ltcfabric.sum} counter for the same launch is
5,944 sectors, so the residue is small in either accounting. The kernel takes
$146.9\,\us$, compared with $144.5\,\us$ for same-die access. This small
difference agrees with the nearly flat L2-resident latency curve in
Figure~\ref{fig:latency}.

Together, these measurements are consistent with remote data being brought into
the accessing die's local L2 and reused on subsequent accesses. The low measured
overhead does not mean that a fabric hop is inexpensive. Rather, once the data
is cached locally, the dependent access chain no longer needs to pay that cost
on every access.

\textbf{Cross-die writes continue to generate fabric traffic.}
An 8\,MiB write produces 33.55M write sectors, nearly doubling the total sector
count from 33.6M to 67.1M. Runtime also increases from 292 to $441\,\us$. In
this workload, repeated reads can reuse cached lines, but writes continue to
generate cross-die traffic. The counter and runtime changes are consistent with
the lower bandwidth measured for cross-die writes.

\textbf{DRAM traffic remains nearly unchanged.}
The four 256\,MiB read configurations each transfer 1.073\,GB from DRAM, with a
spread of 0.0059\% across them; the four 8\,MiB read configurations transfer
about 8.4\,MB, as expected for their working-set size. The four 256\,MiB write
configurations transfer approximately 1.004\,GB.
The performance differences therefore do not come with a corresponding change in
measured DRAM traffic. Instead, the counters point to differences in L2 and
fabric traffic. This supports separating access latency from effective bandwidth
in our model, rather than attributing the results to increased DRAM demand.

These results also explain why the effect of memory placement depends on the
accessor. A local kernel that repeatedly reads resident data can reuse its local
L2 copy even when the original data is placed on the other die. A remote reader
does not necessarily benefit from the same cache relationship. The next
experiment tests whether placing the source data near a remote reader improves
transfer performance.

\subsection{Memory Placement for Remote Accessors}
\label{sec:res:copy}

\textbf{Prediction.}
When a remote GPU reads data from the local GPU, placing the source buffer on
the appropriate die should reduce the cost of the transfer. Compute binding
should have little effect if local SMs do not initiate the relevant accesses.

\textbf{Method.}
We use nvbandwidth to measure transfers initiated by both copy engines and SMs,
covering local and remote access patterns. We test the four placement states and
add a one-sided placement experiment in which only the source or only the
destination buffer is localized. This lets us distinguish the effect of moving
the read source from that of changing the receiving buffer's placement.

\begin{table}[H]
\centering\small
\caption{nvbandwidth in four states, speedup against the same-binary stock baseline (three
repetitions, median). Bold marks the best state in the row.}
\label{tab:nvb}
\begin{tabular}{@{}llrrrr@{}}
\toprule
\textbf{Path} & \textbf{Size} & \textbf{Stock (GB/s)} & \textbf{Memory-only} & \textbf{Compute-only} & \textbf{Full-LD} \\
\midrule
\multirow{3}{*}{Remote read, CE} & 64\,KiB  & 7.4   & \textbf{1.72} & 1.00 & \textbf{1.72} \\
                                 & 1\,MiB & 75.7  & \textbf{2.08} & 1.15 & \textbf{2.08} \\
                                 & 256\,MiB & 744.7 & 1.08 & 1.00 & 1.08 \\
\midrule
\multirow{2}{*}{Remote read, SM} & 1\,MiB & 114.0 & 1.29 & \textbf{1.53} & 1.41 \\
                                 & 4\,MiB & 314.5 & 1.19 & 1.23 & \textbf{1.24} \\
\midrule
Remote write, CE & 1\,MiB & 204.6 & 1.01 & 1.00 & 1.00 \\
Remote write, SM & 1\,MiB & 143.4 & 1.00 & \textbf{1.55} & 1.43 \\
\midrule
\multirow{3}{*}{Local D2D, CE} & 64\,KiB  & 45.5  & 1.00 & 1.15 & \textbf{1.32} \\
                               & 1\,MiB & 663.7 & 1.00 & 1.05 & \textbf{1.18} \\
                               & 256\,MiB & 3236.6 & 0.53 & 0.72 & 0.52 \\
\midrule
Local D2D, SM & 1\,MiB & 599.2 & 1.00 & 1.11 & \textbf{1.23} \\
Local D2D, SM & 256\,MiB & 3136.1 & 0.53 & 1.01 & 0.50 \\
\bottomrule
\end{tabular}
\end{table}

\textbf{Results.}
For a remote copy-engine read of a 1\,MiB buffer, memory placement alone
delivers a $2.08\times$ speedup, while compute binding alone delivers a
$1.15\times$ speedup. The one-sided experiment shows where the larger gain comes
from. Placing only the source buffer yields 157.7\,GB/s, while placing only the
destination yields 75.3\,GB/s, close to the 75.33\,GB/s baseline measured in the
same run.

The stock baseline for this path is not a stable constant. Across independent
processes it is bimodal, spanning 74--101\,GB/s, because the outcome depends on
which die the interleaved pages happen to land on. We quote the low mode
throughout, so the speedups we report are the larger of the two that the same
experiment can produce; pooling all runs would give a baseline near
100.7\,GB/s and correspondingly smaller ratios.

The source-only result supports the prediction that the main benefit comes from
placing the data near the remote reader, rather than changing the receiving
side. It also shows why memory placement and compute binding should be evaluated
separately: the copy engine initiates the transfer, so restricting local SM
execution does not directly control the relevant access.

The other transfer cases show a different pattern. For D2D transfers initiated
by SMs, compute binding is the effective control, while memory placement alone
provides little benefit. For large local copies involving 256\,MiB,
configurations that pin memory achieve only about $0.53\times$ the baseline
bandwidth. This agrees with the streaming measurements: when bandwidth
dominates, restricting memory to one die can cost more than localization saves.

The results therefore depend on which component issues the access. Remote
copy-engine reads benefit primarily from placing the source data near the
reader, whereas SM-initiated transfers respond differently to compute placement.
The one-sided experiment further shows that placing the destination alone does
not explain the remote-read improvement.

\subsection{The Cost of Localized Allocation}
\label{sec:mem:cost}

\textbf{Motivation.}
Localized memory delivers benefits only when its performance gains outweigh the 
overhead of placement setup. VMM allocation and release involve driver-side work
that is not included in the GPU bandwidth measurements above. For workloads that
frequently create and destroy allocations, this overhead can offset the savings
during execution.

\textbf{Method.}
We measure the full create-to-free cycle across seven VMM APIs, with allocation
sizes from 1\,KiB to 1\,GiB and multiple concurrency levels. To locate the
allocation-size threshold, we also sweep the 30--66\,MiB range in 2\,MiB
increments, taking 30 measurements per point. We then break down the 1\,GiB cost
by API and compare cold allocation with allocation from a warmed memory pool.

\begin{figure}[H]
  \centering
  \includegraphics[width=0.68\textwidth]{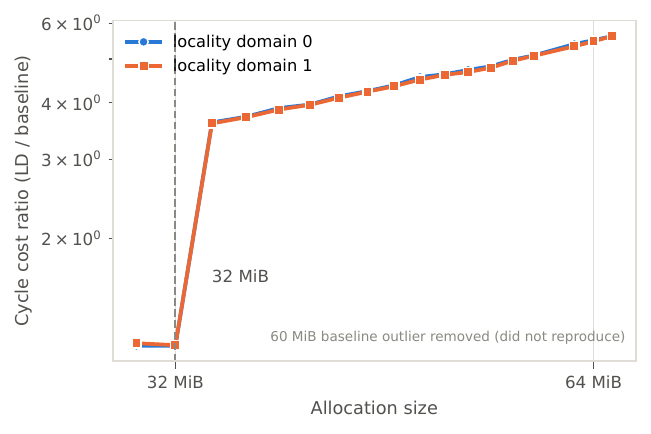}
  \caption{Full create-to-free VMM cycle, LD divided by baseline. 30--66\,MiB is scanned in
  2\,MiB steps with $n=30$ per point; the two domains differ by less than 2\% at every point.}
  \label{fig:vmmstep}
\end{figure}

\begin{table}[H]
\centering\small
\caption{VMM full allocation/free cycle (P50, one GPU, one thread) and the extra cost of LD.
The 60\,MiB baseline point is a known non-reproducing outlier and is excluded.}
\label{tab:vmm}
\begin{tabular}{@{}rrrrl@{}}
\toprule
\textbf{Size} & \textbf{Baseline (\us)} & \textbf{LD (\us)} & \textbf{Ratio} & \textbf{Extra cost} \\
\midrule
2\,MiB   & 123.1  & 144.8   & 1.18$\times$  & $+21.7\,\us$ \\
32\,MiB  & 135.9  & 158.3   & 1.16$\times$  & $+22.4\,\us$ \\
34\,MiB  & 134.9  & 492.0   & \textbf{3.65$\times$} & $+357.1\,\us$ \\
64\,MiB  & 147.9  & 814.0   & 5.50$\times$  & $+666.1\,\us$ \\
512\,MiB & 507.6  & 5876.5  & 11.58$\times$ & $+5368.9\,\us$ \\
1\,GiB   & 877.7  & 11543.4 & \textbf{13.15$\times$} & $+10665.7\,\us$ \\
\bottomrule
\end{tabular}
\end{table}

\textbf{Results.}
The overhead changes sharply around 32\,MiB rather than increasing at a constant
rate with allocation size. From 2\,MiB up to 32\,MiB, the additional cost stays
near $22\,\us$, giving a ratio of 1.16--1.18$\times$ relative to ordinary
allocation. At 34\,MiB, the additional cost jumps to about $357\,\us$ and
thereafter grows by approximately $10\,\us$ per MiB. Both locality domains show
similar behavior, with the transition occurring between 32 and 34\,MiB.

Two caveats qualify the ``below 32\,MiB'' figure. First, it describes requests
of 2\,MiB and larger. Sub-2\,MiB requests are rounded up to the 2\,MiB
granularity and show a larger relative overhead, about $34\,\us$ or
1.27--1.29$\times$, because the same fixed cost is amortized over less memory.
Second, the ratios above are for the full create-to-free cycle; individual APIs
vary more than the cycle does.

We find that the driver's allocation path is a major source of
the overhead through API breakdown. With one thread and a 1\,GiB allocation, \code{cuMemCreate}
increases from 194.0 to $5532.7\,\us$ (28.5$\times$), while \code{cuMemRelease}
increases from 475.4 to $5817.3\,\us$ (12.2$\times$). In the same setting,
\code{cuMemMap}, \code{cuMemSetAccess}, and \code{cuMemAddressReserve} change
relatively little.

Concurrency makes the driver overhead more pronounced. With eight GPUs and eight
threads, \code{cuMemSetAccess} reaches 44.8 times its baseline cost. The 1\,GiB
allocation ratio reaches 12.10$\times$, with a P99 ratio of 17.22$\times$.

Two further experiments help narrow down the cause. Enabling a green context
changes the 1\,GiB result only slightly: 11.54\,ms with a green context versus
11.58\,ms without one. This suggests that green-context use is not responsible
for the allocation overhead. Running both locality domains concurrently also
fails to amortize the cost: the measured time is 80.71\,ms, compared with a
6.64\,ms ordinary baseline.

\begin{figure}[H]
  \centering
  \includegraphics[width=0.64\textwidth]{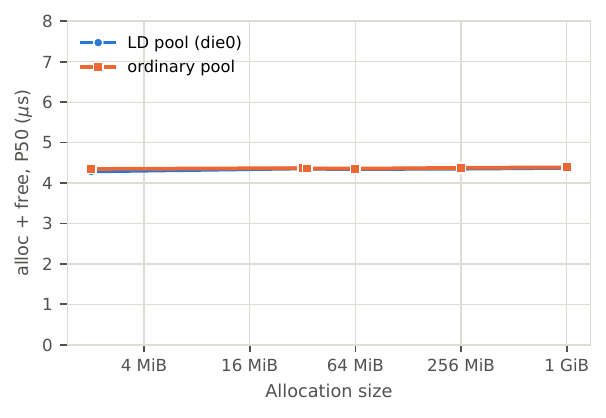}
  \caption{A warmed pool (release threshold $=\infty$) makes
  \code{cuMemAllocFromPoolAsync}+\code{cuMemFreeAsync} indistinguishable between an LD pool and
  an ordinary pool across 2\,MiB--1\,GiB.}
  \label{fig:pool}
\end{figure}

\textbf{Implications.}
The overhead is concentrated in allocation and release, rather than in each
subsequent GPU access. Reusing memory from a warmed pool largely removes the
difference: LD and ordinary pools differ by only $-0.06$ to $-0.01\,\us$ across
sizes from 2\,MiB to 1\,GiB, with zero allocation-ordinal read-back errors. A
pool that returns physical memory on every free, using a threshold of zero,
still incurs a 14--18\% overhead, similar in scale to direct VMM allocation.

These results suggest that localized memory is practical when allocations are
created once and reused. Repeatedly creating, freeing, or trimming localized
allocations can make driver overhead a substantial part of end-to-end execution
time.

\subsection{Tracing and Reducing the Management Overhead}
\label{sec:mem:rootcause}

The sharp increase in allocation cost around 32\,MiB led us to examine the
driver's mapping path. Three observations helped narrow down the cause. First,
the threshold coincides with \path{PMA_LOCALIZED_MEMORY_ALLOC_STRIDE}, which is
32\,MiB and defines the granularity at which PMA guarantees physical contiguity
for localized memory. Second, each call to the mapping routine costs about
$13\,\us$, whether it handles a 2\,MiB range or a much larger one. Third, the
green-context experiment shows that compute placement has little effect on
allocation time.

Together, these observations suggest that the overhead comes mainly from
repeatedly calling the mapping routine, rather than from writing the page-table
entries themselves. We traced the problem to the static BAR1 mapping path and
made two changes to reduce the cost.

\subsubsection{Root cause}

BAR1 is the GPU's second PCIe Base Address Register. It exposes VRAM through a
system address range for CPU and peer-DMA access. During initialization, the
driver builds a static translation table using a mapping of the form
\code{BAR1 VA = VRAM physical address + startOffset}, with a 2\,MiB page size.
This table needs to be updated only for allocations that are not of the default
kind, exceed the default page size, or use localized memory.

The relevant code is \code{_kbusUpdateStaticBar1VAMapping_TU102}. Its
early-exit condition handles ordinary allocations, but localized allocations do
not meet that condition and proceed to the mapping loop. The loop advances by
\code{memdescGetContiguity(...) ? mapSize : pageSize}. Because the memdesc for a
localized allocation is not marked \code{PHYSICALLY_CONTIGUOUS}, the contiguity
check returns false and the loop falls back to a 2\,MiB step.
As a result, mapping a 1\,GiB allocation requires 512 iterations. Each iteration
issues a separate driver call and carries its own \code{DMA_TLB_INVALIDATE}.

The time breakdown explains why this is expensive. Page-table writes account for
only 13--22\% of each call. The remaining 77--87\% is spent on a full-chip
memory barrier (\code{kbusFlush_HAL}), which makes the page-table updates
visible to the GPU, together with TLB invalidation. These operations account for
86.9\% of the cost at 2\,MiB and 77.5\% at 1\,GiB. Since synchronization costs
roughly the same amount per call, repeating it for every 2\,MiB page makes the
total overhead depend heavily on the number of iterations.

\subsubsection{Two complementary changes}

Our fix makes two changes to the mapping path.

The first change combines physically contiguous pages into larger mapping runs.
Instead of assuming that adjacent logical offsets correspond to adjacent
physical addresses, the driver checks each page's physical address and extends a
run only while physical contiguity holds. It then passes the whole run to the
HAL in a single call.

This reduces the number of calls from $O(\text{size}/2\,\mathrm{MiB})$ to
$O(\text{size}/32\,\mathrm{MiB})$. For a 1\,GiB allocation, the count falls from
512 to 32. The physical-address check is necessary because the optimization
bypasses the existing contiguity check; merging pages without verifying their
physical addresses could produce an incorrect mapping.

The second change defers synchronization until the loop finishes. Each mapping
call uses \code{DMA_DEFER_TLB_INVALIDATE}, followed by a single memory barrier
and TLB invalidation after all runs have been processed. This avoids repeating
synchronization for every run.

The two changes address different parts of the cost. Coalescing reduces the
number of mapping calls, while deferred synchronization reduces the number of
barriers and invalidations. The second change offers no benefit for a single 
run below 32\,MiB, since there's no repeated synchronization to eliminate. 
We have submitted the changes to Nvidia's open-gpu-kernel-modules repository
\footnote{https://github.com/NVIDIA/open-gpu-kernel-modules/issues/1499}
\footnote{https://github.com/NVIDIA/open-gpu-kernel-modules/pull/1500}.

\begin{table}[H]
\centering\small
\caption{Due to the static-BAR1 fix: \code{cuMemCreate} P50 in \us{} (one GPU, one thread) under each
patch combination, and the gain of A+B over the unpatched LD path. Patch B alone acts on sync
groups and Patch A alone on call count, so the four columns separate their contributions.
Below 32\,MiB there is a single run and both patches are no-ops.}
\label{tab:vmmfix}
\begin{tabular}{@{}rrrrrrr@{}}
\toprule
\textbf{Size} & \textbf{LD-off} & \textbf{LD-on} & \textbf{B only} & \textbf{A only} & \textbf{A+B} & \textbf{A+B vs LD-on} \\
\midrule
2\,MiB   & 8.9    & 20.4    & 20.3    & 20.6    & 20.4    & 1.00$\times$ \\
32\,MiB  & 13.3   & 26.2    & 26.0    & 25.8    & 26.2    & 1.00$\times$ \\
64\,MiB  & 9.0    & 353.1   & 115.2   & 56.7    & 50.0    & \textbf{7.07$\times$} \\
128\,MiB & 9.8    & 684.7   & 205.6   & 95.5    & 77.8    & 8.80$\times$ \\
256\,MiB & 10.3   & 1359.8  & 388.7   & 178.0   & 132.9   & \textbf{10.23$\times$} \\
512\,MiB & 119.1  & 2850.2  & 914.0   & 485.6   & 395.3   & 7.21$\times$ \\
1\,GiB   & 145.6  & 5745.3  & 1863.4  & 1018.8  & 818.1   & \textbf{7.02$\times$} \\
\bottomrule
\end{tabular}
\end{table}

\subsubsection{Performance gains and remaining overhead}

Table~\ref{tab:vmmfix} shows the effect of these changes on \code{cuMemCreate}.
The improvement is substantial for allocations above 32\,MiB: 7.07$\times$ at
64\,MiB, 10.23$\times$ at 256\,MiB, and 7.02$\times$ at 1\,GiB.

The fix does not remove the entire overhead. At 1\,GiB, localized allocation
still takes $818.1\,\us$, compared with $145.6\,\us$ for the non-localized path,
leaving a 5.6$\times$ gap. This remaining cost is consistent with PMA providing
physically contiguous runs of at most 32\,MiB. Removing that limit would require
changes to PMA, which are outside the scope of this fix.
Below 32\,MiB, the fix has no effect. The mapping is handled as a single run, so
neither combining multiple runs nor deferring synchronization reduces the work.
The roughly $2\times$ overhead for small allocations therefore remains.

We also had to ensure that the optimization preserved the original
synchronization behavior. An earlier version used \code{PTE_UPGRADE} for
deferred invalidation and appeared to improve allocation times at every size,
including below 32\,MiB. That result was misleading because this type also skips
the system memory barrier, so the change did more than batch synchronization.
Using \code{PTE_DOWNGRADE}, the type selected by the original routine, removes
this unintended effect. The apparent improvement for small allocations
disappears, as expected. All results in Table~\ref{tab:vmmfix} use the corrected
implementation.

\textbf{Measurement scope.}
Table~\ref{tab:vmmfix} measures \code{cuMemCreate} alone on a B200 with CUDA
13.4 and driver 615.71.09. Tables~\ref{tab:vmm} and \ref{tab:amortize} report
the full seven-step allocation cycle on a B300. These measurements are not
directly comparable and should not be subtracted from one another; the
difference in scope, not only in platform, accounts for most of the gap. The
mapping mechanism and the 32\,MiB threshold are consistent across the two
platforms, but the reported latencies apply to their respective experimental
setups.

\textbf{Correctness checks and limitations.}
We performed three checks: reading back BAR1 contents page by page and comparing
them with per-page tags written through CUDA; releasing and immediately
reallocating memory to test reuse; and injecting failures into the mapping path.
All three passed.

The BAR1 check is necessary because CUDA virtual addresses are translated
through the GMMU page table and do not exercise the static BAR1 mapping. A
successful CUDA readback alone therefore cannot verify the BAR1 contents. The
first and third checks required driver instrumentation that is absent from the
clean patch, so their results apply strictly to the instrumented build.

The patch also changes error handling. In the original code, a later successful
run could overwrite an earlier error, allowing the caller to receive success
even when part of the requested range had not been mapped. The patch preserves
and returns the first error instead. Neither version rolls back PTEs already
written, so a failed operation can leave residual entries. In the original path,
these entries could be overwritten incidentally during a subsequent release. The
patch makes partial mapping failures explicit, but does not provide rollback.
Adding rollback would require changes to the release path and remains outside
the scope of this fix.

\subsubsection{Implications for allocation and amortization}

The measurements show that much of the overhead for large localized allocations
comes from repeated synchronization in the driver's mapping path, rather than
from locality domains themselves. The fix recovers a substantial part of this
cost above 32\,MiB, but leaves the small-allocation overhead and the residual
cost associated with PMA's contiguity granularity unchanged.

The fix also changes how we should interpret the allocation-amortization results
below. The overhead is still paid when memory is allocated, rather than on every
GPU access, so a warmed pool can amortize it. For large allocations, the fix
also makes cold allocation substantially cheaper. The repayment counts in
Table~\ref{tab:amortize}, which use the original driver path, should therefore
be treated as upper bounds when the fix is applied to allocations above 32\,MiB.
We do not report rescaled counts because the original and patched measurements
were collected on different platforms.

\subsection{Amortization: When Placement Costs Are Repaid}
\label{sec:mem:amortize}

\textbf{Question.}
Localized memory has an upfront allocation cost, while its performance benefit
is earned during subsequent accesses. The relevant question is how many accesses
or transfers are needed to recover that cost.

\textbf{Method.}
We divide the measured additional allocation cost by the measured saving per
access or transfer. We use only cases in which the experiments in
Section~\ref{sec:res:copy} identify memory placement as the effective control.
An earlier version of this analysis used the saving from a collective, even
though that improvement came from compute binding rather than memory placement.
We exclude that case because it would overstate the benefit attributable to
localized memory.

\begin{table}[H]
\centering\small
\caption{Number of accesses needed to repay one LD allocation. Basis: a cross-die HBM dependent
access saves \SI{193.4}{ns}; one remote CE read of 1\,MiB saves $7.18\,\us$ (75.7$\to$157.2\,GB/s).}
\label{tab:amortize}
\begin{tabular}{@{}rrrr@{}}
\toprule
\textbf{Allocation} & \textbf{Extra cost (\us)} & \textbf{Repaid by dependent accesses} & \textbf{Repaid by 1\,MiB CE reads} \\
\midrule
2\,MiB   & 21.7    & 112      & 4 \\
32\,MiB  & 22.4    & 116      & 4 \\
64\,MiB  & 666.1   & 3,444   & 93 \\
512\,MiB & 5,368.9 & 27,758 & 748 \\
1\,GiB   & 10,665.7 & 55,143 & 1,486 \\
\bottomrule
\end{tabular}
\end{table}

\textbf{Results.}
The results of three different cases are shown in Table~\ref{tab:amortize}.
For a 2\,MiB buffer repeatedly read by a
remote engine, the measured allocation cost is recovered after four transfers.
For a 1\,GiB allocation, the reported cost of 10.7\,ms is equivalent to 55,143
dependent cross-die accesses. Recovering this cost therefore requires sufficient
reuse over the allocation's lifetime. For a large streaming workload, placement
does not repay its cost through access savings because localization itself
reduces bandwidth. Here, the bandwidth penalty outweighs the potential latency
benefit.

The transfer-based counts inherit the baseline ambiguity described in
Section~\ref{sec:res:copy}: they use the 75.7\,GB/s low mode. Had we used the
pooled 100.7\,GB/s baseline instead, the saving per 1\,MiB transfer would fall
from $7.18\,\us$ to roughly $3.7\,\us$ and the counts would roughly double. The
access-based counts do not depend on that choice, because they use the
dependent-load latency increment rather than a bandwidth ratio.

These results give a practical way to decide when to use LDs. Frequently reused
buffers accessed by a remote reader can benefit from localization. Large
allocations need enough reuse to recover their setup cost. Bandwidth-bound
streaming workloads, by contrast, should generally retain an interleaved layout
when it provides access to more memory bandwidth.

The amortization counts above use the original, unpatched management path. With
the static-BAR1 fix, the additional allocation cost for sizes above 32\,MiB
decreases by the factors reported in Table~\ref{tab:vmmfix}, so the
corresponding repayment counts should also decrease. Counts below 32\,MiB remain
unchanged because the fix does not modify that path. We retain the original
measured counts rather than reporting rescaled estimates, since the original and
patched measurements were collected on different platforms and rescaling would
imply more precision than the data supports.

\section{Compute Placement}
\label{sec:compute}

\subsection{Motivation and Experimental Challenge}

The third case is a local kernel that repeatedly reads data already resident in
memory, such as the weights of a GEMM. Section~\ref{sec:mem:counter} showed that
cross-die reads can be served from the accessor's local L2 after the first
remote access. Repeated reads may therefore pay little additional cost for
crossing the fabric. However, when a kernel runs on SMs across both dies, its
data cannot be local to every SM at the same time. Local-L2 reuse reduces
repeated fabric traffic, but it does not make a split execution equivalent to
one where data and compute are placed on the same die.

Our hypothesis is that memory placement and compute placement work best
together. Memory placement alone leaves some executing SMs remote from the data,
while compute placement alone leaves some data remote from the executing SMs.
Enabling both should avoid these two problems. Testing this hypothesis is not
straightforward, because the two controls also change resource availability:
memory placement restricts where data resides, and compute placement restricts
the available SMs. We use four placement states and a mismatch control to
separate these effects.

\subsection{Method}

\textbf{Workload.}
We use DeepGEMM's masked grouped GEMM, which lets us control where the data
resides and where the computation runs. A mixture-of-experts layer issues eight
independent group GEMMs, which we divide evenly between the two dies. The
weights stay fixed across iterations, so we can place them once and reuse them.
We focus on small batch sizes, where execution time is short and placement
effects are easier to observe. All variants run inside a single CUDA graph and
use prepacked weight scales, so graph construction and input preparation are
outside the measured interval.

\textbf{Placement states.}
We evaluate the four states listed in Table~\ref{tab:states}. The compute-only
state uses the same 70-SM resource count and memory footprint as the
full-placement state, but does not explicitly place the weights on the execution
die. Comparing these two states shows what memory placement adds when compute is
already restricted to one die.

\textbf{Mismatch control.}
The mismatch state places the weights on one die and runs the kernel on the
other. The workload, kernel, and SM count remain unchanged; only the data and
execution locations are deliberately misaligned. Comparing mismatch with full
placement shows the cost of this misalignment. It may also change fabric traffic
and contention, so we examine branch timings and concurrency before attributing
the slowdown to a particular mechanism.

\textbf{Mechanism instrumentation.}
We record CUDA events on both execution streams without inserting a profiler
into the measured loop. This lets us check whether the branches still overlap
and measure each branch's execution span. Profiler traces substantially perturb
the overlap. We also found that a five-state counter collection came from a
reference GEMM rather than DeepGEMM. We therefore do not use either the profiler
trace or those counters to explain the measured slowdown.

\textbf{Boundary conditions.}
We also flush 256\,MB of L2 before each kernel and repeat the experiments with
FP8 and FP4. These tests show whether the gains persist when the weights are not
already resident in cache and when the amount of data transferred changes.

\subsection{Results}

\begin{table}[H]
\centering\small
\caption{DeepGEMM under the r2 protocol (eight groups split 4+4 across dies, one CUDA graph,
prepacked weight scales). Hot wall clock in \us, median of three repetitions; speedups are
relative to the same-protocol stock.}
\label{tab:dg}
\begin{tabular}{@{}llrrrrr@{}}
\toprule
\textbf{Precision} & \textbf{M} & \textbf{Stock} & \textbf{Memory-only} & \textbf{Compute-only} & \textbf{Full-LD} & \textbf{Full speedup} \\
\midrule
\multirow{4}{*}{FP8}
 & 1  & 12.14 & 11.79 (1.03$\times$) & 11.72 (1.04$\times$) & 11.52 (1.05$\times$) & 1.05$\times$ \\
 & 2  & 29.36 & 39.08 (0.75$\times$) & 28.41 (1.03$\times$) & \textbf{21.46 (1.37$\times$)} & 1.37$\times$ \\
 & 8  & 29.62 & 39.46 (0.75$\times$) & 28.39 (1.04$\times$) & \textbf{21.64 (1.37$\times$)} & 1.37$\times$ \\
 & 16 & 29.41 & 38.91 (0.76$\times$) & 27.90 (1.05$\times$) & \textbf{21.41 (1.37$\times$)} & 1.37$\times$ \\
\midrule
\multirow{2}{*}{FP4}
 & 8  & 15.84 & 19.04 (0.83$\times$) & 15.20 (1.04$\times$) & 14.94 (1.06$\times$) & 1.06$\times$ \\
 & 16 & 16.06 & 18.74 (0.86$\times$) & 15.23 (1.05$\times$) & 14.89 (1.08$\times$) & 1.08$\times$ \\
\bottomrule
\end{tabular}
\end{table}

\begin{figure}[H]
  \centering
  \includegraphics[width=0.72\textwidth]{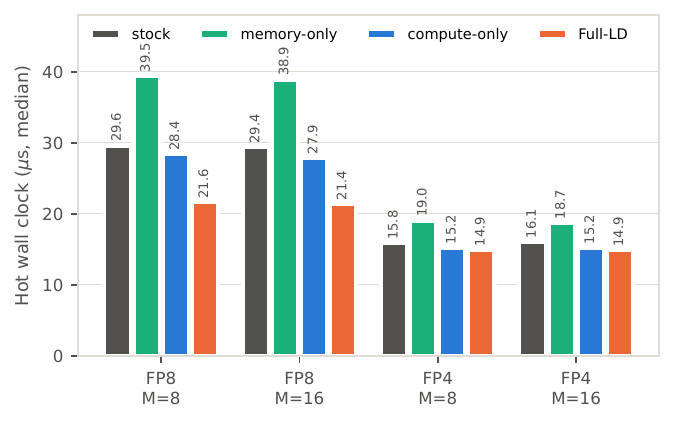}
  \caption{DeepGEMM masked hot wall clock. Memory placement alone is a net loss
  ($0.75$--$0.86\times$), compute binding alone is nearly neutral ($1.03$--$1.05\times$), and
  only the combination reaches $1.37\times$.}
  \label{fig:dg}
\end{figure}

The four states behave quite differently. For FP8, with memory placement alone,
performance falls by about 25\% at every $M\geq 2$, to $0.75$--$0.76\times$ the
stock performance; at $M=8$ runtime increases
from $29.62$ to $39.46\,\us$, roughly a third longer. The weights reside on one
die, but execution still uses all 148
SMs. SMs on the other die must therefore access remote weights, even if repeated
reads can later hit in their local L2.

Compute placement alone is close to the stock configuration, at $1.04\times$.
The kernel runs on one die, but some of the weights remain remote. With both
controls enabled, runtime drops from $29.62$ to $21.64\,\us$, a $1.37\times$
speedup.

The gain largely disappears when we flush L2 before each kernel. Under
cold-cache conditions, speedup ranges from $1.01$ to $1.05\times$ for FP8 and
from $0.94$ to $1.05\times$ for FP4. Even in the hot-data case, FP4 achieves
only $1.06$--$1.08\times$. The benefit is therefore not a general improvement
from restricting execution to one die. It depends on how much data can be reused
and whether the saved access latency is large enough to offset the loss of
bandwidth and other fixed costs.

\subsection{Interaction Between Placement Controls}

Memory placement alone achieves $0.75\times$ the stock performance, and compute
placement alone achieves $1.04\times$. With both enabled, performance reaches
$1.37\times$. The combined result is much better than either control on its own.

For comparison, if the two controls had independent multiplicative effects,
their combined speedup would be $0.75\times 1.04=0.78\times$. The measured
speedup of $1.37\times$ is about $1.76\times$ this value. The two controls
therefore interact strongly rather than contributing independent gains.

The reason is straightforward: memory placement alone does not move all
execution closer to the weights, and compute placement alone does not move all
the weights closer to execution. Applying both aligns the data with the SMs that
use it, while still allowing the workload to use both dies.

The single-domain results show why this alignment matters. Pinning the entire
GPU to one domain, without distributing work across both dies, reaches only
$0.63\times$ the stock performance. This configuration keeps compute and
data local, but also gives up SMs and memory-channel capacity. Locality alone is
not enough to make up for the lost parallelism.

\subsection{The Cost of Misaligned Placement}

The mismatch control increases runtime by $1.74\times$, from $21.39$ to
$37.22\,\us$. Because the kernel, workload, and SM count are unchanged, the
difference cannot be explained by a change in these factors. It does not,
however, identify a single physical cause: misalignment can also change fabric
traffic and contention. We examine the branch timings and overlap to narrow down
the explanation.




\textbf{Both branches take longer when they run together.}

\begin{figure}[H]
  \centering
  \includegraphics[width=0.66\textwidth]{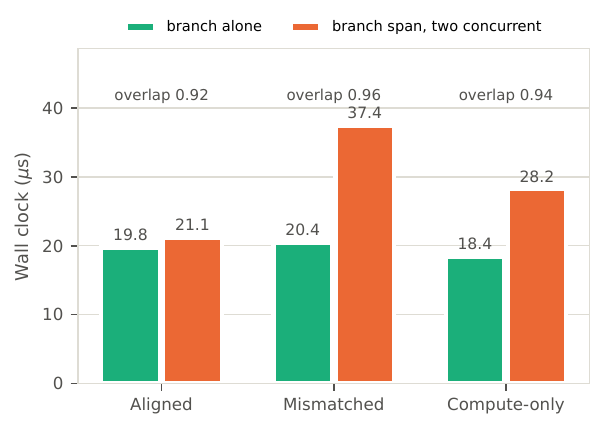}
  \caption{Event-instrumented branches (median of three independent processes). Bars are the mean
  span of the two branches; the label above each pair is their interval overlap. Under mismatch
  the overlap does not fall; each branch simply becomes slower.}
  \label{fig:g5b}
\end{figure}

As shown in Figure~\ref{fig:g5b}, under mismatch, 
both branch spans increase from about $21\,\us$ to $37.4\,\us$.
When each branch runs alone, however, its runtime changes only from
$19.5/20.0\,\us$ to $20.2/20.6\,\us$, an increase of roughly 2--4\%. Remote
weight access has only a small effect on an isolated branch, but a much larger
effect when both branches run concurrently.

This points to contention when both branches access weights across dies. The
independent primitive measurements found that cross-die read bandwidth is 73\%
of same-die bandwidth (Section~\ref{sec:mem:access}), which provides a
plausible explanation for the slowdown. The branch timings also make reduced
concurrency an unlikely primary cause. However, without valid DeepGEMM fabric
counters, we cannot directly confirm that the remote-access path is saturated.

\subsection{Sensitivity to Data Residency}
\label{sec:res:boundaries}

The placement benefit changes substantially when the weights are not resident in
L2. After flushing 256\,MB of L2 before each kernel, speedup falls to
$1.01$--$1.05\times$ for FP8 and $0.94$--$1.05\times$ for FP4. The gain seen
with resident weights therefore does not carry over to the cold-cache case.

When the working set exceeds the effective L2 capacity, more accesses reach
DRAM. In this regime, restricting memory to one die also limits the bandwidth
available to the kernel, making it harder for the latency savings to compensate
for the bandwidth loss. The cold-cache results put a clear limit on the benefit
of placement: it helps most when the kernel can reuse resident data.

\subsection{Decode Attention Under Bandwidth Pressure}
\label{sec:res:attn}

\textbf{Workload and experimental setup.}
We next examine decode attention, which repeatedly scans historical K/V data.
Unlike grouped GEMM with resident weights, this workload continuously moves
through a large sequence and puts more pressure on memory bandwidth.

We implement a chunked decode-attention kernel in CUDA C++. Each decode CTA
scans part of the K/V sequence, computes a partial attention result, and writes
an intermediate softmax state and output. A separate merge kernel combines these
intermediate results.

Our initial GQA implementation uses 64 query heads and eight KV heads, but shows
less than 1\% sensitivity to placement at a logical K/V scan rate of about
265\,GB/s. We therefore build a higher-throughput scan workload with 64
independent output rows per request. Each row owns a separate bf16 K/V sequence
with head dimension 128. For batch size $B$, sequence length $S$, and KV tile
size $t$, the decode grid contains $64B\lceil S/t\rceil$ CTAs.

The stock configuration divides the output rows evenly between two ordinary CUDA
streams, uses interleaved K/V allocation, and has access to all 148 SMs. The
LD-aware configuration keeps the same row partition, kernel, and decode grid,
but binds the two branches to separate 70-SM Green Contexts and places their K/V
and intermediate buffers on the corresponding dies.

We evaluate 30 combinations of sequence length and batch size on B300. Each CTA
uses eight warps, and the runs use warm-cache, pipelined execution. We warm up
each configuration for five iterations and measure 20 iterations, excluding
compilation and initialization. Both implementations use the same input data,
arithmetic, and correctness checks.

\begin{table}[H]
\centering\small
\caption{Best-versus-best comparison at the eight points whose default-tile gain exceeded
$1.20\times$; the decision rule was frozen before the run. Each form chooses the tile that minimises
its own median time (in parentheses), so the two columns are not matched on tile size: $R$ is best
versus best, not an equal-tile ratio. Classes use the pre-registered
boundaries: SURVIVES $R>1.05$, SHRINKS $0.95\le R\le1.05$, REVERSES $R<0.95$. Hot wall clock in
ms, median of three repetitions.}
\label{tab:attn}
\begin{tabular}{@{}rrrrrl@{}}
\toprule
\textbf{S} & \textbf{B} & \textbf{Baseline best (tile)} & \textbf{Row split best (tile)} & \textbf{$R$} & \textbf{Class} \\
\midrule
2048  & 2 & 0.0362 (128) & 0.0412 (128) & 0.879 & REVERSES \\
2048  & 4 & 0.0680 (256) & 0.0725 (512) & 0.938 & REVERSES \\
4096  & 1 & 0.0366 (128) & 0.0422 (128) & 0.867 & REVERSES \\
4096  & 2 & 0.0681 (256) & 0.0729 (512) & 0.935 & REVERSES \\
4096  & 4 & 0.1304 (512) & 0.1361 (512) & 0.958 & SHRINKS \\
8192  & 1 & 0.0684 (256) & 0.0732 (512) & 0.934 & REVERSES \\
8192  & 2 & 0.1304 (512) & 0.1365 (512) & 0.955 & SHRINKS \\
16384 & 1 & 0.1309 (512) & 0.1379 (1024) & 0.949 & REVERSES \\
\bottomrule
\end{tabular}
\end{table}

\begin{figure*}[t]
  \centering
  \includegraphics[width=\textwidth]{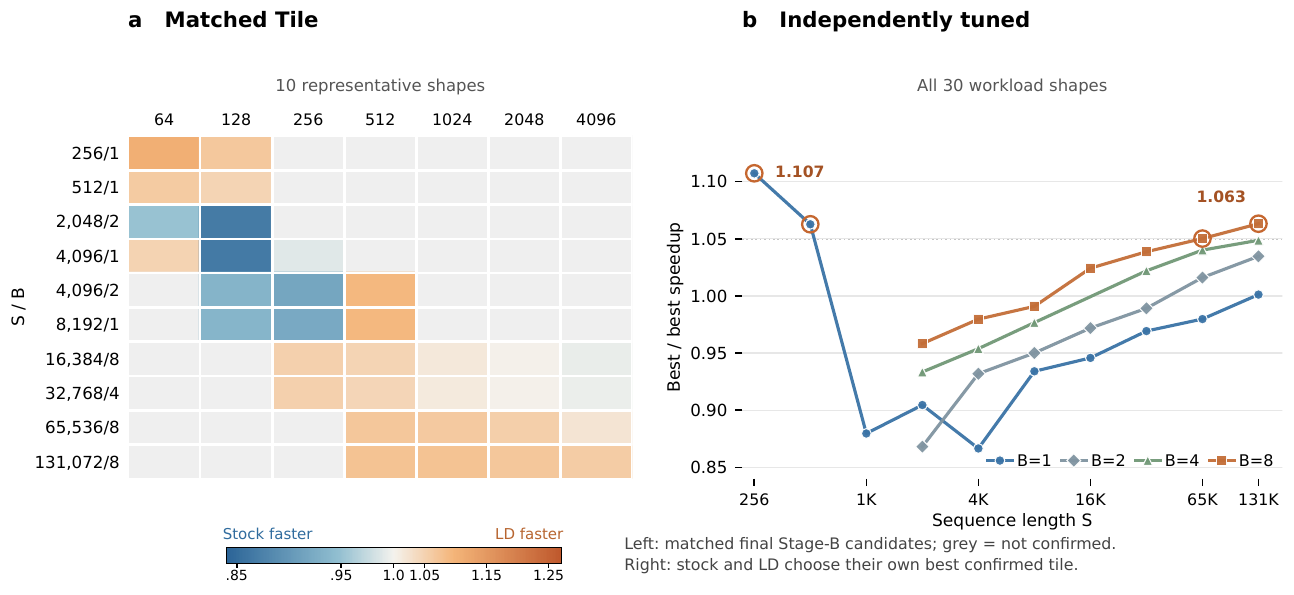}
  \caption{Decode attention under matched and independently tuned KV tiles.
  (a) Same-tile speedup for ten representative workload shapes from the
  confirmed candidate set; gray cells were not selected for final confirmation.
  (b) Best-vs-best speedup across all 30 workload shapes, allowing stock and
  LD-aware execution to select their fastest measured tile independently. Values
  above unity favor LD-aware execution.}
  \label{fig:attn:tile-matrix}
\end{figure*}

\textbf{Fixed-configuration performance.}
We first hold the sequence length $S$, batch size $B$, and KV tile size $t$
constant, and compare stock with LD-aware execution.
Figure~\ref{fig:attn:tile-matrix}(a) shows same-tile speedups for representative
workload shapes and tile sizes.

With the original default tile, LD-aware execution is more than 5\% faster on 15
of the 30 workloads, including eight with gains above 20\%. The maximum speedup
is $1.27\times$. The largest gains occur when the stock decode grid exposes
relatively little parallel work, with approximately 0.9--3.7 decode CTAs per SM
at the strongest gain points.

Changing the tile size can change the result substantially. Across the
matched-tile sweep, LD-aware execution generally does better with coarser KV
tiles, while finer tiles often favor stock. Intermediate-sized workloads are
particularly sensitive to this choice. For example, at $S=4096$ and $B=1$,
LD-aware execution reaches $1.051\times$ with tile 64 but falls to $0.867\times$
with tile 128. With the original tile size of 2048, the same workload reaches
$1.27\times$. The placement configuration has not changed; changing the amount
of K/V work assigned to each CTA is enough to turn a large gain into a slowdown.

The four placement states help explain some of these differences. For long
sequences, memory placement and compute binding can work together. At
$S=131072$, $B=8$, and $t=2048$, memory placement alone reaches $0.793\times$
stock performance, and compute binding alone reaches $0.936\times$. Together,
they reach $1.063\times$.

Short sequences are also sensitive to the intermediate buffers. At $S=256$ and
$S=512$, localizing the partial/output allocation bundle improves the already
aligned compute-and-K/V configuration by another 3.6--3.8\%. The gain is
negligible for the longest sequence. Since this bundle also changes allocator
details, we cannot attribute the difference to physical locality alone. The
results show that the useful placement choices vary with workload shape.

\textbf{Independent tile tuning.}
A same-tile comparison does not necessarily show how much benefit placement can
deliver after tuning. The best tile for stock need not be the best tile for
LD-aware execution. Tile size determines how much K/V data each CTA scans, how
many independent CTAs the grid exposes, and how much work the merge kernel must
do. Since the two configurations use different compute resources and memory
paths, they may favor different tile sizes. In our measurements, their best
tiles differ for 19 of the 30 workload shapes.

We therefore compare the best measured configuration of each variant for every
workload. We first test all legal power-of-two tiles from 64 to 4096, then
select leading candidates independently for stock and LD-aware execution. We
remeasure the union of these candidates in ten fresh processes per workload,
balancing the two execution orders. For each variant, we choose the tile with
the lowest median latency across the repeated processes and compare the two
selected configurations.

\textbf{Best-vs-best results.}
Figure~\ref{fig:attn:tile-matrix}(b) shows the results after independent tuning.
Only four workload shapes retain more than 5\% speedup with LD-aware execution:
$S=256$, $B=1$ ($1.107\times$); $S=512$, $B=1$ ($1.063\times$); $S=65536$, $B=8$
($1.050\times$); and $S=131072$, $B=8$ ($1.063\times$).

The remaining gains are concentrated in two regions: short sequences with small
batches, and long sequences with large batches. For the 16 intermediate-sequence
workloads ($S=1024$--16384), median speedup is $0.948\times$, and none exceeds
$1.05\times$. For the longest sequence, LD-aware execution fares better as batch
size increases: at $S=131072$, speedup rises from $1.001\times$ at $B=1$ to
$1.063\times$ at $B=8$.

\begin{figure}[H]
  \centering
  \includegraphics[width=0.98\textwidth]{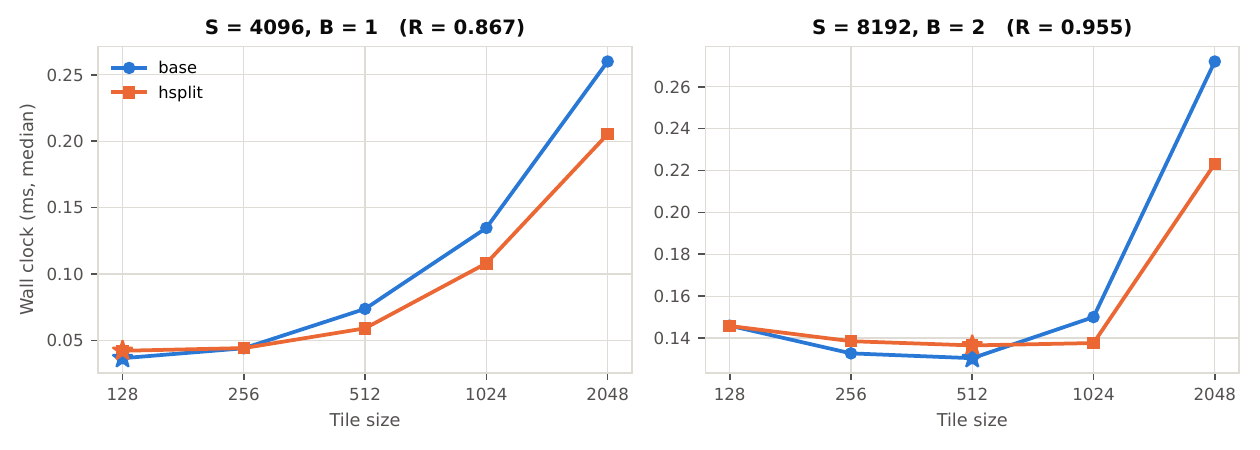}
  \caption{Tile sweep at two of the eight pre-registered points; stars mark each form's own best
  tile. The baseline's time falls faster as the tile shrinks, so across the eight points it gains
  $2.07$--$7.11\times$ from choosing its own tile against $1.59$--$4.87\times$ for the row split.
  The ordering measured at the default tile therefore reverses once both forms are independently tuned.}
  \label{fig:attn-sweep}
\end{figure}

The workloads with the largest gains under the default tile do not remain the
strongest beneficiaries after tuning. None of the eight workloads that initially
showed more than $1.20\times$ speedup retains a positive best-vs-best result.

Consider $S=4096$, $B=1$. Reducing the tile from 2048 to the best measured
setting cuts stock latency from 260.0 to $36.6\,\us$. LD-aware latency also
falls, from 205.3 to $42.2\,\us$, but the initial $1.27\times$ advantage becomes
a ratio of $0.867\times$. Finer K/V partitioning helps stock much more in this
case, removing the apparent benefit of placement.

\textbf{Understanding the remaining gains.}
A KV-mismatch control helps show what die alignment changes. Deliberately
placing each branch's K/V on the opposite die increases long-sequence decode
latency by about 32\%. The penalty is also present when branches run
individually and is concentrated in the K/V-scanning stage. Keeping K/V data on
the same die as the branch that scans it therefore avoids a substantial
remote-access cost.

That does not mean the aligned configuration must outperform stock by the same
amount. Stock already uses the memory resources of both dies and retains more
freedom in how computation is scheduled. For short sequences, K/V scanning takes
less time, so the configuration of partial and output buffers becomes more
visible in the overall result.

\textbf{Implications.}
Grouped GEMM and decode attention respond differently to placement. With
resident weights, grouped GEMM reaches $1.37\times$ speedup when both controls
are enabled. For decode attention, after independent tile tuning, gains above
5\% remain for only four of the 30 workload shapes, with speedups of
$1.05$--$1.11\times$.

Tile tuning gives decode attention another way to improve performance: changing
the K/V chunk size can expose more CTA parallelism without restricting execution
to individual dies. The benefit of locality must therefore be compared with what
ordinary kernel tuning can already achieve, as well as with the SMs and memory
bandwidth lost through partitioning. In a serving system, scheduling and kernel
selection introduce further choices, so die-aware placement is one part of the
overall optimization rather than a standalone fix.

Three findings emerge from these experiments. First, memory placement and
compute placement can reinforce each other: neither control alone delivers the
full grouped-GEMM gain. Second, the useful configuration depends on workload
shape. Short decode sequences are more sensitive to auxiliary-buffer placement,
while long sequences benefit from aligned K/V access at sufficiently large batch
sizes. Third, independent tile tuning substantially changes the results and
removes the largest apparent gains under the default configuration. What matters
is not just whether data and compute are placed together, but whether that
placement still helps after accounting for available resources and kernel
tuning.

\subsection{Summary}

For grouped GEMM with resident weights, memory and compute placement work best
together. Neither control alone delivers the full gain, while the combined
configuration reaches $1.37\times$ speedup. The mismatch control also shows that
separating data and execution placement can be costly even when the branches
continue to overlap. Branch timings point to contention during concurrent remote
weight access, although the lack of valid DeepGEMM fabric counters means we
cannot directly confirm the mechanism.

The cold-cache GEMM tests and decode-attention experiments show where this
benefit ends. Placement gains depend on data residency, memory bandwidth, and
the parallelism exposed by the kernel configuration. For decode attention, the
$1.27\times$ speedup under the default tile disappears after independent tile
tuning for the workloads that initially showed the largest gains. Locality
should therefore be evaluated together with resource availability and kernel
tuning, rather than assumed to improve performance in every configuration.

\section{Communication Placement}
\label{sec:comm}

\subsection{Attributing the Benefit of Compute Placement}

The second accessor class is a local SM that issues a transfer, as in a
collective whose TMA engine and participating threads run on the same GPU. Our
hypothesis is that compute placement matters most for this case, while memory
placement alone may do little or even hurt when it reduces available bandwidth.
There is, however, an important confounding factor: the configurations that bind
compute to one die use 70 SMs instead of 148, whereas memory placement alone
leaves all 148 SMs available. A collective might run faster with fewer SMs for reasons
unrelated to which die they occupy.

We therefore use a separate control to distinguish SM count from placement. We
first measure the gain and how it changes with message size, then compare LD
against MPS configurations with a similar number of SMs but execution spanning
both dies. Finally, we examine the execution timings to narrow down the possible
mechanism. We also test flash-ar, a second collective workload in which the
critical path involves remote writes and local polling. Comparing these two
workloads lets us test whether the useful placement control depends on the
access pattern, rather than on communication as a whole.

\subsection{Method}

\textbf{General workload.}
We use nccl-tests with symmetric-memory registration to exercise the same
collective path across 2, 4, and 8 ranks, with message sizes from 1\,KB to
64\,MB. We test the four placement states using both All-Reduce and All-Gather.
Our main case is 2-rank All-Gather at 1\,MB. In the reference implementation,
this operation includes an out-of-place self-copy. The hypothesis is that
compute placement makes this copy cheaper by keeping the issuing SMs and the
relevant L2 resources on the same die.

\textbf{SM-count control.}
To separate placement from SM count, we compare the 70-SM compute-only state
with MPS configurations that expose a similar number of SMs while allowing
execution to span both dies. MPS at 50\% and 25\% provides the comparison
points. We set the decision rule before running the experiment: a speedup of at
least $1.25\times$ under MPS would support the explanation that fewer SMs are
enough to produce the gain. A result near $1.0\times$ would instead suggest that
confining execution to one die matters.

\textbf{Launch geometry.}
We also run the same binary in the stock and compute-only states under a tracing
profiler and compare the collective kernel's grid and block dimensions. If these
dimensions are unchanged, the gain cannot be explained by a change in launch
geometry, although this comparison does not rule out every possible difference
in code generation.

\textbf{Second workload.}
Finally, we measure customized all-reduce and all-gather kernels, named flash-ar
and flash-ag, at world sizes of 2, 4, and 8. These
implementations have distinct algorithmic phases that we can selectively remove.
By measuring the affected phase separately, we can determine whether placement
changes the communication itself or another part of the collective.

\subsection{Results}

\begin{table}[H]
\centering\small
\caption{NCCL 2-rank R2 All-Gather in four states (\us, median of three repetitions). Bold marks
the best state per row.}
\label{tab:nccl}
\begin{tabular}{@{}rrrrrr@{}}
\toprule
\textbf{Size} & \textbf{Stock} & \textbf{Memory-only} & \textbf{Compute-only} & \textbf{Full-LD} & \textbf{Compute speedup} \\
\midrule
1\,KB    & 2.93  & 3.05  & 2.93  & 3.05  & 1.00$\times$ \\
64\,KB   & 9.98  & 10.56 & 9.99  & 10.57 & 1.00$\times$ \\
1\,MB    & 15.52 & 15.96 & \textbf{10.78} & 11.32 & \textbf{1.44$\times$} \\
4\,MB    & 19.57 & 19.49 & \textbf{14.80} & 15.60 & \textbf{1.32$\times$} \\
16\,MB   & 30.63 & 30.35 & \textbf{25.78} & 26.43 & \textbf{1.19$\times$} \\
64\,MB   & 75.46 & 74.96 & \textbf{70.30} & 71.99 & 1.07$\times$ \\
\bottomrule
\end{tabular}
\end{table}

\begin{figure}[H]
  \centering
  \includegraphics[width=0.64\textwidth]{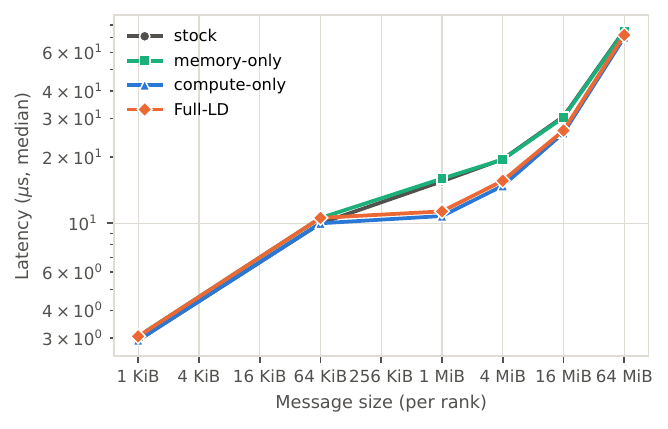}
  \caption{2-rank R2 All-Gather in four states. Memory-only coincides with stock; compute-only
  alone captures the gain and is slightly better than full placement.}
  \label{fig:nccl}
\end{figure}

For 2-rank symmetric-memory All-Gather at 1\,MB, stock takes $15.52\,\us$.
Memory placement alone takes $15.96\,\us$ ($0.97\times$), compute placement
alone takes $10.78\,\us$ ($1.44\times$), and full placement takes $11.32\,\us$
($1.37\times$). In this case, moving the data alone does not help; restricting
execution to one die does.

This configuration is sensitive to run-to-run variation in the stock baseline.
The SM-count control in Section~\ref{sec:res:smcount} repeats the same point on
a different session and obtains $15.76\,\us$ for stock and $10.54\,\us$ for
compute-only, or $1.50\times$. We quote $1.44\times$ here because it comes from
the same session as the other three states, and $1.495\times$ there because it
comes from the same session as the MPS comparison. Both are measurements of the
same configuration; they differ by about 4\% in ratio.

Memory placement can also make a collective slower. For 8-rank R0 All-Reduce at
16\,MB, performance falls to $0.86\times$, with runtime increasing from $85.47$
to $99.86\,\us$. Restricting data to one die limits the bandwidth available to
the operation, and the cost outweighs any benefit from locality.

The compute-only gain becomes smaller as the message grows: $1.44\times$ at
1\,MB, $1.32\times$ at 4\,MB, $1.19\times$ at 16\,MB, and $1.07\times$ at
64\,MB. A separate five-round sweep of the full-placement state over
512\,KB--16\,MB gives an average absolute saving of $4.52\,\us$ with a standard
deviation of $0.24\,\us$. Recomputed from the four-state runs above, the
compute-only saving over the same sizes is $4.88\,\us$; the two agree that the
absolute saving is roughly flat with message size. The absolute saving stays
nearly constant while the relative
speedup shrinks. This trend is consistent with
Equation~\eqref{eq:delta}: a roughly fixed per-operation benefit becomes a smaller fraction
of total runtime as the transfer time grows.

\subsection{Separating Placement from SM Count}
\label{sec:res:smcount}

\begin{table}[H]
\centering\small
\caption{SM-count control: equalising the available SM count while changing die ownership
(NCCL 2-rank R2 All-Gather,
median of three repetitions).}
\label{tab:g1}
\footnotesize
\begin{tabular}{@{}rrrrrr@{}}
\toprule
\textbf{Size} & \textbf{Stock} & \makecell{\textbf{Compute-only}\\\textbf{one die, 70 SMs}} & \makecell{\textbf{MPS 50\%}\\\textbf{both dies}} & \makecell{\textbf{MPS 25\%}\\\textbf{both dies}} & \makecell{\textbf{Compute}\\\textbf{speedup}} \\
\midrule
512\,KB & 16.43 & 11.19 & 16.43 & 16.70 & \textbf{1.468$\times$} \\
1\,MB   & 15.76 & 10.54 & 15.77 & 15.81 & \textbf{1.495$\times$} \\
4\,MB   & 18.84 & 13.91 & 18.92 & 18.77 & \textbf{1.354$\times$} \\
16\,MB  & 30.01 & 24.97 & 30.04 & 30.25 & \textbf{1.202$\times$} \\
\bottomrule
\end{tabular}
\end{table}

\begin{figure}[H]
  \centering
  \includegraphics[width=0.66\textwidth]{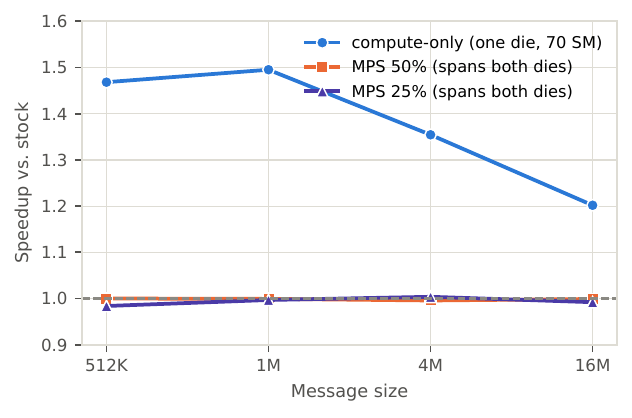}
  \caption{MPS~50\% and compute-only expose a similar number of SMs (about 88 versus 70), but
  MPS spans both dies while compute-only sits in one, and the outcomes are $0.999\times$ versus
  $1.495\times$.}
  \label{fig:smcount}
\end{figure}

The MPS control does not reproduce the LD speedup. At 1\,MB, MPS at 50\% reaches
$0.999\times$, and MPS at 25\% reaches $0.997\times$. Compute-only execution on
one die's 70 SMs reaches $1.495\times$. Reducing the available SM count 
therefore gives essentially no improvement when execution still spans both dies.


The launch-geometry comparison points in the same direction. Both states launch
the same kernel with identical geometry: 16 or 8 blocks of 512 threads each, in
a cluster of four blocks. Only the die on which those CTAs execute differs. A
profiler trace of both states gives a median per-launch kernel duration of
$20.26\,\us$ for stock and $13.52\,\us$ for compute-only. (This pair is a
kernel-duration measurement, not the end-to-end collective latency of
Table~\ref{tab:g1}.) The gain is not explained by a change in grid size or block
dimensions.

With SM count and launch geometry controlled, the main remaining difference is
where those CTAs execute. In the compute-only state, they run on one die and
use that die's L2 slices and scheduling resources. In the stock state, execution
spans both dies and their interconnect. The measurements show that confining the
collective's CTAs to one die improves performance, but they do not tell us
exactly how much comes from L2 behavior, scheduling, or fabric traffic. We could
not obtain usable kernel-level counters for this path
(Section~\ref{sec:res:evidence}).


\subsection{Memory Placement for Communication Control Data}
\label{sec:res:flashar}

The previous experiment shows that compute placement helps an SM-issued
collective. Flash-ar provides a different case. Its slabs, mailboxes, and
counters are managed by the backend and can be placed locally, while the user
payload remains striped across the dies.

\begin{table}[H]
\centering\small
\caption{flash-ar / flash-ag small-message latency gains (P50, relative to the same-binary
baseline; values parsed from the recorded benchmark output).}
\label{tab:flashar}
\footnotesize
\begin{tabular}{@{}lrrr@{}}
\toprule
\textbf{Operation} & \textbf{world 2} & \textbf{world 4} & \textbf{world 8} \\
\midrule
All-Reduce (16\,B--64\,KB) & $-10$ to $-14\%$ & $-6$ to $-10\%$ & $-3$ to $-5\%$ \\
All-Gather ($\leq$128\,KB) & $-11$ to $-21\%$ & $-5$ to $-11\%$ & $+0.4$ to $-4\%$ \\
\midrule
\multicolumn{4}{@{}l}{\emph{Mechanism ablation (world 2, all-gather, 32\,KB)}} \\
CTX=0 (memory only) vs.\ CTX=1 & \multicolumn{3}{l}{$<0.2\%$ difference at 9 of 12 large-message points; max 1.5\%} \\
Spin-wait phase & \multicolumn{3}{l}{1.18 $\to$ 0.78\,\us{} (w2); 1.41 $\to$ 1.34\,\us{} (w8)} \\
Pure transport phase & \multicolumn{3}{l}{send-only / recv-local change $\pm 0.05\,\us$} \\
\bottomrule
\end{tabular}
\end{table}

At world size 2, placing only the slabs and mailboxes locally captures the
entire gain to within 1\% of the configuration that also binds compute. The
phase-deletion experiment on the all-gather variant at 32\,KB shows where the
gain comes from: the spin-wait phase drops from $1.18$ to $0.78\,\us$, while
pure transport changes by no more than $\pm0.05\,\us$. Larger sizes behave the
same way; the 64\,KB and 128\,KB points give $1.23 \to 0.79$ and
$1.27 \to 0.80\,\us$.

The localized objects are the mailbox and counter that a remote rank writes and
the local rank polls. In other words, the critical path includes a remote write
followed by local polling. Moving these control objects closer to the polling
side helps, even though the user payload remains striped. This matches the
remote-access pattern discussed in Section~\ref{sec:res:copy}.

The result changes for large messages. Beyond 32\,MB, localization causes a
1--7\% slowdown. In the ablation, localizing memory while keeping all SMs
available produces results within 0.2\% of those obtained by localizing memory
and binding compute at most sizes, with the largest divergence at world size 2
and 32\,MB, where the memory-only variant is 1.5\% slower. Both are slower than
baseline. The slowdown therefore
follows memory placement, not the reduction in SM count. At these sizes, the
loss of bandwidth from restricting data to one die outweighs the benefit for the
control objects.

\textbf{Interpretation.}
The two collectives respond to different controls. The SM-issued All-Gather
benefits mainly from compute placement, while flash-ar benefits from placing the
backend objects used in remote writes and local polling. Neither result supports
a blanket rule that one placement control is always best for communication. The
useful control depends on which access lies on the critical path.

\subsection{Asymmetry Between Locality Domains}
\label{sec:res:symmetry}

\begin{table}[H]
\centering\small
\caption{flash-ar All-Reduce at world 2, split by GPU pair (64\,KB point, \us). die0's
advantage is concentrated in one pair; die1 is slightly slower than stock everywhere.}
\label{tab:sym}
\begin{tabular}{@{}lrrrr@{}}
\toprule
\textbf{GPU pair} & \textbf{Stock} & \textbf{die0} & \textbf{die1} & \textbf{die0 speedup} \\
\midrule
GPU 0/1 & 3.15 & \textbf{2.68} & 3.22 & \textbf{1.18$\times$} \\
GPU 2/3 & 3.17 & 2.95 & 3.24 & 1.07$\times$ \\
GPU 4/5 & 3.17 & 2.93 & 3.23 & 1.08$\times$ \\
GPU 6/7 & 3.14 & 2.94 & 3.19 & 1.07$\times$ \\
\bottomrule
\end{tabular}
\end{table}

Repeating the primitive matrix on all
eight GPUs gives a cross-die latency increment of 191.4--193.3\,ns when averaged
over the two accessor directions per GPU, a spread of
only 1.9\,ns. The additional latency for L2-resident accesses ranges from 2.7 to
3.3\,ns, and cross-die streaming bandwidth falls by about 27\% from the same-die
value in either direction. At this level, neither die is consistently faster.

The collective results are less symmetric. On GPU pair 0/1, choosing die0 gives
$1.18\times$ speedup at 64\,KB, while the other three pairs reach only
$1.07$--$1.08\times$. Die1 is slower than stock on every pair, averaging
$0.95$--$0.98\times$ over the four sizes per pair; individual small-message
points dip to $0.92\times$. Reversing the run order and
repeating the measurements preserves both patterns, making warm-up and run-order
effects less likely explanations.

Since the primitive measurements are symmetric to within about 1\,ns, the
collective difference is unlikely to come from a large intrinsic latency
difference between the dies. One possibility is that the world-size-2 rank
mapping gives the two dies different roles in the collective, including
different uses of their L2 slices. We could not verify this explanation without
the kernel-level counters unavailable in our environment, so it remains a
hypothesis.

The practical implication is that there is no hardware-based rule to always choose
one die. Die0 is a reasonable starting point for the setup we measured, but
deployments should benchmark both choices with their actual rank mapping and
workload.

\subsection{Summary}

For SM-issued collectives, compute placement is the effective control in our
measurements. The speedup reaches $1.495\times$, compared with $0.999\times$ for
an MPS configuration with a similar SM count that still spans both dies. The
explicit resource-count control is important: a comparison between stock and LD
alone would not show whether the gain came from placement or from using fewer
SMs. The gain also shrinks as message size grows, while the absolute saving
stays nearly constant over the earlier sweep.

Memory placement helps a different part of communication. In flash-ar, the gain
comes from localizing the mailbox and counter involved in remote writes and
local polling, not from placing the user payload locally. For large messages,
the bandwidth loss can outweigh this benefit. Together, the two workloads show
that the useful placement control depends on the access on the critical path,
not simply on whether the workload is a collective.

Finally, the primitive measurements show no consistent preference between the
two dies, even though the collective results differ by die. The difference is
more likely related to the collective's mapping than to a large physical latency
gap between the dies, but we cannot confirm the mechanism with the available
counters.

\section{Discussion: Locality Domains and Alternative GPU-Sharing Mechanisms}
\label{sec:discussion}

\subsection{Placement, Capacity, and Isolation}

Sections~\ref{sec:mem}--\ref{sec:comm} measured both the benefits of locality
domains (LDs) and the costs of using them. We now compare LD with other ways
to share a GPU. The main question is whether these alternatives can provide
the same performance benefit, and what they give up in doing so.

The mechanisms control different aspects of the GPU. MPS limits the resources
available to a process~\cite{nvidia_mps}, while MIG provides hardware
partitioning and isolation~\cite{nvidia_mig}; both have been evaluated for
spatial co-execution~\cite{villarrubia2026comprehensive}. MLOPart exposes
per-die partition devices~\cite{nvidia_mlopart}. Spatial GPU-sharing systems
such as MuxFlow~\cite{zhao2023muxflow}, ParvaGPU~\cite{lee2024parvagpu}, and
VUDA~\cite{xu2026vuda} divide capacity among concurrent jobs without
controlling placement. LD instead controls where memory resides and where
computation runs. These capabilities overlap, but they are
not interchangeable: limiting the number of SMs does not necessarily put the
remaining SMs on the desired die, and isolating a workload does not by itself
align its data with its execution.

\begin{table}[H]
\newcommand{\tagM}{\textcolor{accent}{\textbf{M}}\ }
\newcommand{\tagD}{\textcolor{muted}{\textbf{D}}\ }
\newcommand{\tagI}{\textcolor{muted}{\textbf{I}}\ }
\newcommand{\tagN}{\textcolor{warn}{\textbf{N}}\ }
\centering\footnotesize
\caption{Capability matrix. ``SM split'' asks whether SMs can be partitioned. Evidence tags:
\textcolor{accent}{\textbf{M}} measured, \textcolor{muted}{\textbf{D}} documented,
\textcolor{muted}{\textbf{I}} inferred, \textcolor{warn}{\textbf{N}} not measurable.}
\label{tab:matrix}
\setlength{\tabcolsep}{3.5pt}
\begin{tabularx}{\textwidth}{@{}>{\raggedright\arraybackslash}p{2.1cm}*{5}{>{\raggedright\arraybackslash}X}>{\raggedright\arraybackslash}p{2.0cm}c@{}}
\toprule
\textbf{Mechanism} & \makecell{\textbf{SM}\\\textbf{split}} & \makecell{\textbf{Pick}\\\textbf{die}} & \makecell{\textbf{Memory}\\\textbf{by die}} & \makecell{\textbf{Pool}\\\textbf{shrinks}} & \makecell{\textbf{Change}\\\textbf{at run}} & \textbf{Isolation cost} & \textbf{Source} \\
\midrule
Stock & no & no & no & --- & --- & 1.00$\times$ & \tagM \\
MIG & \tagD yes & \tagD yes & \tagI with instance & yes (89\,GB) & no & \tagN & \tagD \\
MPS & \tagM quota & \tagM \textbf{no} & \tagM yes & yes (SM halved) & \tagM yes & \tagM 0.893$\times$ & \tagM \\
green context & \tagM exact & \tagM yes & no & yes (70/148) & no & \tagM 1.021$\times$ & \tagM \\
memory-only & no & no & \tagM yes & no & --- & \tagM 0.740$\times$ & \tagM \\
MLOPart & \tagM yes & \tagM yes & \tagI with device & yes & no & \tagM 1.23--2.13$\times$ & \tagM \\
native Full-LD & \tagM yes & \tagM yes & \tagM yes & yes (70/148) & no & \tagM 1.367$\times$ & \tagM \\
\bottomrule
\end{tabularx}
\end{table}

The experiments below compare these mechanisms along three dimensions:
which placement choices they expose, how they affect the available resources,
and what overhead they introduce. This distinction matters because the best
configuration in our measurements is not always the one with the fewest active
SMs or the strongest partitioning.

\subsection{MPS: Controlling Capacity Without Controlling Placement}

MPS is the most widely deployed alternative we evaluate. It works in our
environment: the daemon starts, clients attach, and SM limits take effect.
However, it does not provide the same control over placement as LD.

\textbf{MPS does not select a die explicitly.}
The SMs exposed by MPS form a contiguous range starting at SM~0, but this
range spans both dies. Across settings from 25\% to 100\%, the observed
distribution remains approximately 55:45 across the two dies. The driver's
reported SM count also differs from the number that can actually be scheduled.
At the 50\% setting, for example, the driver reports 74 SMs, while 88 are
observed to be schedulable. MPS therefore behaves more like a resource quota
than an exact partition of the SMs on a particular die.

\textbf{Reducing the SM count does not reproduce the LD gain.}
For grouped GEMM, MPS at 50\% takes $32.51\,\us$, achieving $0.893\times$
the stock performance. Full placement without MPS takes $21.22\,\us$ and
achieves $1.367\times$. Combining MPS with full placement reaches only
$0.911\times$ the stock performance.

Restricting the resources also affects memory throughput. For a 256\,MiB
buffer, read bandwidth falls from 6947 to 6253\,GB/s, while write bandwidth
falls from 6414 to 4164\,GB/s. MPS can limit how much of the GPU a process
uses, but that limit does not guarantee that data and execution are placed
together. In these experiments, reducing capacity alone does not recover
the full LD benefit.

\textbf{MPS shares unused capacity rather than enforcing a fixed split.}
We also run two processes performing $4096^3$ bf16 GEMMs, with one process
assigned one quarter of the work. Stock execution takes 592\,ms, compared
with 573\,ms under MPS at 50/50. The shorter job does not have to wait for
the longer job to finish before unused resources can be used by the other
client. This behavior helps utilization under skewed loads, but it is
different from reserving a fixed share of the GPU for each workload.
Multiplexing complementary workloads over the same GPU has similarly been used
to improve inference goodput at the serving layer~\cite{chen2026towards}.

This distinction helps explain the collective results in
Section~\ref{sec:res:smcount}. MPS reduces the available resource count
without explicitly selecting a die, but does not reproduce the collective
speedup. Fewer SMs alone are therefore not enough to explain the benefit
of compute placement.

\subsection{MIG: Isolation Does Not Guarantee the Same Performance}
\label{sec:res:mig}

MIG is the most relevant alternative when hard isolation is required.
Performance isolation for co-located inference processes has been studied in
edge GPU systems~\cite{martin2026performance}. Vendor
documentation~\cite{nvidia_mig} describes a Blackwell profile with two
instances, each exposing 70 SMs and 89\,GB of memory, matching the measured
$2\times70+8$ SM structure. We could not run MIG directly because it is
disabled in our container. Instead, we test the performance assumption
behind a proposed MIG-based bound using a substitute configuration. The
results below should not be read as direct MIG measurements.

Our substitute is deliberately favorable to the partitioned configuration.
We place four groups on one die to emulate a single instance, then
estimate the time for two such instances to run concurrently under a
zero-interference assumption. This assumption is what the experiment tests.

\begin{table}[H]
\centering\small
\caption{The MIG-style split, measured and bounded (DeepGEMM FP8 masked M=8, hot, \us, three
repetitions). The $1.43\times$ figure assumes two instances with zero interference.}
\label{tab:mig}
\begin{tabular}{@{}lrr@{}}
\toprule
\textbf{Configuration} & \textbf{Wall clock (\us)} & \textbf{vs.\ full stock} \\
\midrule
Stock: 148 SMs, eight groups & 29.00 & 1.00$\times$ \\
One die (domain0 / domain1), four groups & 20.22 / 20.44 & 1.43$\times$ (upper bound) \\
One die, eight groups (uneven load) & 46.25 & \textbf{0.63$\times$} \\
Same process, 4+4, compute-only & 28.44 & \textbf{1.02$\times$} \\
Same process, 4+4, memory + compute (Full-LD) & 21.14 & \textbf{1.37$\times$} \\
\bottomrule
\end{tabular}
\end{table}

The estimated runtime is $20.33\,\us$ per instance, the average of the two
single-die runs in Table~\ref{tab:mig}, compared with
$29.00\,\us$ for the full GPU, giving a nominal $1.43\times$ speedup.
But the corresponding measured configuration, which uses the same 4+4
split in one process with compute binding but without memory placement,
takes $28.44\,\us$, or $1.02\times$ the stock performance.

The difference comes from concurrent execution. A single branch of this
compute-only configuration runs in $18.1$--$18.6\,\us$; under the aligned
full-placement configuration of Section~\ref{sec:compute} the isolated branches
take $19.5$ and $20.0\,\us$. Either way an isolated branch is far cheaper than
the $28.44\,\us$ that two concurrent branches take. The measured configuration
therefore does not behave as though the two branches run independently.
This is consistent with interference when data and execution are not aligned,
although the substitute is not a direct measurement of MIG.

The second column of Table~\ref{tab:mig} shows another cost of partitioning.
Running all eight groups on one die takes $46.25\,\us$, or $2.27\times$
the $20.33\,\us$ required by four groups. Reducing the available SMs without
reducing the workload accordingly reverses the apparent benefit.

The trade-off is different for the two mechanisms. MIG provides hardware
isolation, but a workload must fit the resources assigned to its instance.
LD allows memory and compute placement to be adjusted without necessarily
shrinking the resource pool, although it does not provide the same isolation
as a dedicated hardware partition.

These measurements do not establish that MIG itself achieves $1.02\times$,
nor that its performance matches our substitute. They show a narrower point:
the nominal $1.43\times$ estimate depends on the assumption that concurrent
instances incur no interference. The measured substitute does not support
that assumption.

\subsection{MLOPart: Per-Die Placement and Its Management Cost}

MLOPart exposes per-die partition devices~\cite{nvidia_mlopart}, so it can
select placement in a way that MPS cannot. Its measured cost is mainly in
driver-side memory management.

In prior measurements, VMM allocation and release for a single partition
cost $0.98$--$1.28\times$ the native-LD baseline at the P50 median. With
two concurrent partitions, the ratios increase to $1.23$--$2.13\times$.
P90 medians range from $1.27$ to $3.47\times$, with individual points
reaching $4.89$--$6.54\times$. With four threads per partition, we also
observed \code{CUDA_ERROR_INVALID_DEVICE} and sticky errors.

These measurements come from a prior study, not a rerun in our current
experimental environment, so the numerical comparison should be treated
with care. They nevertheless show that providing per-die placement is only
part of the problem: the cost and reliability of managing those partitions
also matter. MLOPart and LD can express overlapping placement choices, but
their operating costs are not the same.

\subsection{Summary}
\label{sec:res:synthesis}

\begin{figure}[t]
  \centering
  \includegraphics[width=0.76\textwidth]{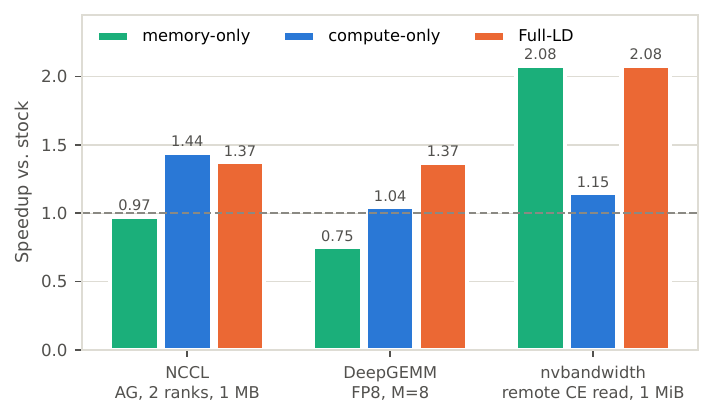}
  \caption{Three states over three accessor shapes, all relative to the same-binary stock.
  NCCL gains most from compute-only and DeepGEMM from Full-LD; the remote copy-engine read gains most from
  memory-only or Full-LD; only DeepGEMM needs both, and its $1.37\times$ exceeds the product of the two
  single states.}
  \label{fig:fourstates}
\end{figure}

Figure~\ref{fig:fourstates} summarizes the main finding: no single placement
state is best for every accessor. The useful control depends on which part of
the access path is expensive. The observations below summarize the results and findings.

\begin{enumerate}[leftmargin=1.7em,itemsep=0.35em]

\item \textbf{Cross-die access has different costs for different operations.}
\evMeas{}

A dependent HBM access incurs an additional $193.4\,\mathrm{ns}$ when it
crosses dies, measured with the accessor on die~0; the reverse direction gives
$191.5\,\mathrm{ns}$ and the two-direction average is $192.5\,\mathrm{ns}$.
Averaged over both directions per GPU, the value agrees to within about
$2\,\mathrm{ns}$ across eight GPUs. For L2-resident reads, the measured
increment is only about $3\,\mathrm{ns}$. Cross-die writes still incur a
substantial penalty
when data is L2-resident. For 256\,MB streaming reads, the cross-die
bandwidth is about 73\% of the same-die figure, a 27\% reduction, but only
about 51\% of the interleaved baseline, a reduction of about 49\%. The two
references measure different things: the same-die figure uses one die's HBM
channels, whereas the interleaved figure lets a single-die accessor draw on
both dies' channels, so it exceeds the same-die number as well.

These results show that the cost depends on the operation and where its
data is served from. The exact numbers may differ on other platforms, but
the measurements here do not support treating every cross-die access as
having the same cost.

\item \textbf{Locality can reduce latency while also reducing bandwidth.}
\evMeas{}

For the measured streaming workload, a single-die layout reaches
3628\,GB/s for reads and 3219\,GB/s for writes, compared with 5230 and
3795\,GB/s for the striped layout. Restricting memory to one die can
therefore reduce throughput even when it improves locality.

Decode attention shows why this trade-off matters in a real kernel. With
the default tile, LD-aware execution reaches up to $1.27\times$ speedup.
After stock and LD-aware configurations are tuned independently, none of
the eight workloads with default-tile speedups above $1.20\times$ retains
a positive best-vs-best result. At the continuously batched decode
operating point, with 14.6 blocks per SM, the default-tile ratio is already
$0.984$--$1.003\times$.

A locality comparison must therefore account for the kernel configuration
and the available resources. Moving data closer is not enough if the
resulting layout limits bandwidth or if ordinary kernel tuning removes the
original performance gap.

\item \textbf{The useful placement control depends on the accessor.}
\evMeas{}

A remote reader benefits from memory placement alone, reaching
$2.08\times$ the baseline for a 1\,MiB transfer. SM-issued collectives
benefit mainly from compute binding, reaching $1.495\times$, while memory
placement is neutral or harmful in the measured cases. For resident-data
GEMM, neither control alone delivers the full gain; enabling both reaches
$1.37\times$.

The results support a simple rule: place data near the side that accesses
it, and place compute near the data when the executing SMs issue repeated
accesses. A controlled experiment showing that the opposite control
consistently wins for the same accessor, with other factors held fixed,
would challenge this interpretation.

\item \textbf{The collective speedup is not explained by fewer SMs alone.}
\evMeas{}

For the measured collective, one-die compute placement reaches
$1.495\times$, while an MPS configuration with a comparable SM count
but execution spanning both dies reaches $0.999\times$. Both states launch the
same kernel with identical geometry---16 or 8 blocks of 512 threads in a
four-block cluster---so only the die on which the CTAs execute differs. The
comparison argues
against reduced SM count or changed launch geometry as the explanation
for the speedup.

This conclusion would need to be reconsidered if a comparable resource
count spanning both dies reproduced the same gain.

\item \textbf{Memory and compute placement reinforce each other for resident-data GEMM.}
\evMeas{}

For the measured GEMM, memory placement alone achieves $0.75\times$
the stock performance, compute placement alone achieves $1.04\times$,
and full placement reaches $1.37\times$. If their effects were independent
and multiplicative, the expected combined speedup would be
$0.75\times1.04=0.78\times$. The measured result is approximately
$1.76\times$ this baseline.

This comparison shows a strong positive interaction for this workload.
It does not imply that combining locality controls will have the same
effect for every kernel.

\item \textbf{The mismatch slowdown appears when branches run concurrently.}
\evMeas{} with a mechanism caveat.

Under mismatch, branch overlap remains high at 0.957, while each branch's
execution span grows from about $21\,\us$ to $37.4\,\us$. When a branch
runs alone, its runtime increases by only 2--4\%. The slowdown is therefore
consistent with contention when concurrent branches access remote weights,
rather than simply a loss of concurrency.

However, we did not obtain valid DeepGEMM fabric counters, so the
measurements do not establish a unique physical cause.

\item \textbf{LD--VMM management overhead changes sharply with allocation size.}
\evMeas{}

With the original allocation path, the measured VMM cycle adds about
$22\,\us$ of overhead from 2\,MiB up to 32\,MiB, then rises sharply between 32 and
34\,MiB and grows by about $10\,\us$ per MiB above it. Requests below 2\,MiB
are rounded up to the 2\,MiB granularity and pay a larger relative overhead,
about 1.27--1.29$\times$. Most of the
additional cost is in object creation and release. A warmed allocation
pool removes the measured difference in the tested configurations,
showing that reusing allocations can amortize much of this overhead.

This does not mean every allocation path is free of overhead, nor that
pooling eliminates all management costs.

\item \textbf{The primitive measurements show no consistent preference for either die.}
\evMeas{}

Across eight GPUs, the measured cross-die latency increment varies by
only $1.9\,\mathrm{ns}$. The collective results nevertheless show a
recurring difference between die choices for particular GPU pairs.
The source of this difference remains unknown; the current evidence
does not establish an intrinsic physical advantage for either die.

\item \textbf{The MIG-style speedup estimate depends on a zero-interference assumption.}
\evMeas{} for the substitute; \evDoc{} for documented MIG capabilities.

The nominal $1.43\times$ estimate falls to $1.02\times$ in the measured
substitute configuration, where concurrent branches do not behave as
independent workloads. This challenges the estimate under the measured
execution conditions, but it is not a direct measurement of MIG performance.

\item \textbf{Capacity, isolation, and placement are different controls.}
\evMeas{} + \evDoc{}

MPS primarily controls resource quotas, while MIG provides hardware
partitioning and isolation. Neither reproduces arbitrary data--execution
placement in the experiments considered here. MLOPart can express per-die
placement, but its measured concurrent management costs are substantial.
LD exposes memory and compute placement without necessarily reducing
the resource pool, but does not provide the same isolation as MIG.

These mechanisms should therefore be compared by the control they provide,
the resources they restrict, and the overhead they introduce, rather than
treated as interchangeable ways to divide a GPU.

\end{enumerate}

\subsection{Limitations of the Mechanism Analysis}
\label{sec:res:evidence}

Two parts of the analysis have weaker evidence than the performance results
themselves. These limitations restrict how precisely we can explain the
mechanisms, but they do not remove the measured performance differences.

\textbf{We could not obtain kernel-level counters for the collective path.}
The profiler's kernel-replay mode fails with a memory-save error (rc=9),
while application replay times out (rc=124) against NCCL's memory management.
We did not pursue an injection-based alternative because it requires
rebuilding the plugin configuration, and the container has no network access.

As a result, the attribution in Section~\ref{sec:res:smcount} rests on the
SM-count control and launch-geometry comparison. The explanation for why
the self-copy becomes cheaper under compute placement remains an inference,
not a mechanism directly confirmed by kernel-level counters.

\textbf{One counter dataset was attributed to the wrong kernel.}
The five-state counter rows were later found to belong to a reference GEMM
(\code{nvjet_sm103_*}, grid 1312), not DeepGEMM's kernel (grid 70).
The recorded values have not changed, but they cannot be used to explain
DeepGEMM's behavior. The conclusions in Section~\ref{sec:compute} therefore
rely on event-level timings and concurrency measurements, not on those
counter rows.

The distinction is important throughout this discussion: we directly measure
the performance changes and how they depend on placement, workload, and
resource configuration. For some results, we can narrow down the likely
explanation, but we cannot yet identify a single hardware mechanism with
certainty.
\section{Conclusion}
\label{sec:conclusion}

This paper studies how CUDA Locality Domains affect GPU workloads on
Blackwell B300, focusing on which placement control reduces the
cross-die cost on a workload's critical access path. We use a
two-term cost model to distinguish the latency saved by avoiding
cross-die accesses from the bandwidth lost through localized memory
placement, and derive predictions based on the accessor.

The measurements show that these two costs affect workloads differently.
A dependent HBM access incurs a cross-die latency increment of
\SI{193.4}{ns} with the accessor on die~0 (\SI{192.5}{ns} averaged over both
directions), while the increment for L2-resident reads is only
about \SI{3}{ns}. Cross-die streaming bandwidth is approximately
27\% lower than the same-die baseline and approximately 49\% lower
than the interleaved baseline. The read/write asymmetry
is consistent with remote reads being backfilled into the accessor's
local L2, while writes continue to traverse the fabric; DRAM traffic
remains unchanged in the measured comparison. The effective placement
control also depends on the accessor. Memory placement alone improves
remote reads of a 1\,MiB buffer by $2.08\times$. For SM-issued
collectives, compute binding achieves $1.495\times$ the stock
performance, compared with $0.999\times$ for a comparable SM count
spanning both dies. For resident-data GEMM, neither control is
sufficient on its own: memory placement reaches $0.75\times$ stock
performance, compute placement reaches $1.04\times$, and combining
both reaches $1.37\times$. Decode attention provides a contrasting
case: the apparent $1.27\times$ gain with a fixed default tile
largely disappears once the stock and LD-aware variants are tuned
independently. None of the eight workloads with default-tile gains
above $1.20\times$ keeps a positive result, and only four of the 30
shapes retain more than a $5\%$ speedup. These results support
accessor-based predictions, but do not imply uniform behavior across
all workloads in a class.

The experiments also expose two practical complications. LD--VMM
management overhead rises sharply around 32--34\,MiB rather than
acting as a uniform per-allocation cost. A warmed allocation pool
can amortize this overhead when memory is reused, but does not remove
it from every allocation path. In addition, partitioning work into
one single-die instance per die does not guarantee the benefit
predicted by a zero-interference model. Our substitute experiment
challenges the assumption behind the nominal $1.43\times$ MIG-style
bound, but does not measure MIG directly.

The main lesson is that placement should follow the accessor's
critical path, not a general expectation that locality is beneficial.
Its value depends on the cross-die accesses it removes, the bandwidth
it sacrifices, and the compute and memory-system resources left
available to the workload.

\begin{table}[H]
\centering\footnotesize
\caption{Switch selection by accessor and data volume.}
\label{tab:guide}
\begin{tabularx}{\textwidth}{@{}>{\raggedright\arraybackslash}p{4.6cm}>{\raggedright\arraybackslash}p{4.2cm}>{\raggedright\arraybackslash}X@{}}
\toprule
\textbf{Accessor / shape} & \textbf{Recommendation} & \textbf{Evidence} \\
\midrule
CE or peer GPU reading this GPU's data (small to medium, $\le$4\,MiB) &
Place memory, do not bind compute &
Remote CE read at 1\,MiB: memory-only $2.08\times$, compute-only $1.15\times$ \\
Flag or counter written remotely and polled locally &
Place memory only &
flash-ar at world 2: CTX=0 captures the full gain (within 1\%) \\
Communication issued by local SMs (small messages, TMA path) &
Bind compute; this also holds for large messages &
NCCL 1\,MB compute-only $1.44$--$1.50\times$; 8-rank AR 16\,MB compute-only
$1.33\times$ vs.\ full $0.94\times$ \\
Local SMs repeatedly reading resident data (GEMM weights) &
Place memory and bind compute to the same die, and split work by die &
DeepGEMM $1.37\times$; one die running all eight groups $0.63\times$ \\
Any large streaming access ($>$ L2 per die) &
Do not enable LD; bind compute only if needed &
Local D2D 256\,MiB: memory-only $0.53\times$, full $0.52\times$ \\
Bandwidth-bound kernel (decode attention) &
Do not enable; a gain at a default tile does not survive independent tile tuning &
Best-vs-best comparison with each form on its own best tile: 0/8 survive,
$R=0.867$--$0.958$; a fixed-default-tile baseline shows $1.27\times$ \\
Allocating on a hot path &
Build a localized pool at startup, allocate and free inside it, never trim &
Warmed pool LD vs.\ ordinary $<0.1\,\us$; cold pool still 14--18\% slower \\
Objects localised once ($\le$32\,MiB) &
Direct VMM allocation is fine; it adds about 22\,\us{} of overhead &
A 2\,MiB buffer repays itself in four CE reads \\
Large objects ($>$32\,MiB) on a driver with the static-BAR1 fix &
Object creation is 7--10$\times$ cheaper than unpatched, but still 5.6$\times$ the non-localized
path; pre-allocate, do not allocate per request &
1\,GiB \code{cuMemCreate} 5745.3 $\to$ 818.1\,\us; residual gap is PMA's 32\,MiB run limit \\
Small objects ($\le$32\,MiB), patched or not &
Unchanged by the fix; treat exactly as before &
Fix is a no-op below 32\,MiB (single run) \\
Multi-tenant or multi-process isolation &
MIG on bare metal, or MLOPart; do not expect MPS to control die assignment &
MPS spans both dies at every setting; MIG is documented \\
Which die to choose &
Default to die0, but measure per deployment &
Hardware is symmetric; GPU pair 0/1 shows a clear die0 gain, others 7\%, die1 slightly slower \\
\bottomrule
\end{tabularx}
\end{table}

\subsection{Practical Guidance}

Table~\ref{tab:guide} summarizes how to choose between memory and
compute placement. The decision starts with the accessor that pays
the cross-die cost, then considers whether that cost is large enough
to outweigh the bandwidth and resource trade-offs.

\textbf{Choose placement based on the accessor, not the workload name.}
Workloads with similar names can respond to different controls.
SM-issued collectives benefit primarily from compute binding, whereas
a collective whose critical path includes a remotely written,
locally polled mailbox can benefit from memory placement. Application
categories alone are therefore not a reliable guide to which control
will help.

\textbf{Do not assume that localized memory provides more bandwidth.}
For large streaming workloads, localizing memory can reduce throughput
relative to striped placement. When accesses generate substantial
DRAM traffic, the bandwidth penalty can outweigh the reduction in
cross-die latency. Striped memory may then be preferable, while
compute binding should be evaluated separately against the loss of
available SMs and memory-system parallelism. Working-set size alone
does not determine which configuration will be faster.

\textbf{Verify placement and benchmark the target configuration.}
An allocation API should not be assumed to guarantee a particular
die placement. In our remote-read benchmark, uncontrolled placement
of \code{cudaMalloc} buffers produces a bimodal baseline of
74--101\,GB/s, making results sensitive to the actual placement.
Likewise, the primitive measurements show no consistent intrinsic
preference between dies, yet some collective configurations retain
a performance asymmetry. Deployments sensitive to die choice should
verify allocation and rank placement, then benchmark the actual
workload rather than assume that one die is always preferable.

\subsection{Limitations}

\begin{enumerate}[leftmargin=1.7em,itemsep=0.3em]

\item \textbf{One hardware platform.}
The model's two platform-specific inputs---the cross-die latency
increment of approximately \SI{193}{ns} and the relative bandwidth---
were measured on eight B300 GPUs with CUDA 13.4. Other Blackwell
configurations may differ in die count, fabric topology, and driver
behavior. Applying the model elsewhere requires measuring these
quantities on the target platform.

\item \textbf{No kernel-level counters for the collective path.}
Both profiler replay modes fail against NCCL's memory management
in our environment, and we did not attempt the injection-based
alternative. The attribution of the communication gain therefore
rests on the SM-count control and launch-geometry comparison.
Why the self-copy becomes cheaper remains an interpretation of
the placement results, not a mechanism directly verified by
kernel-level counters.

\item \textbf{One counter dataset was misattributed.}
The five-state kernel-counter dataset belongs to a reference GEMM,
not the DeepGEMM kernel under study. The recorded values themselves
are unchanged, but they cannot support claims about DeepGEMM's
execution. Our conclusions for that kernel instead rely on
event-level timing and concurrency measurements.

\item \textbf{MIG was not measured directly.}
The container does not permit us to instantiate MIG. Whether a
MIG instance's SM set matches a locality domain, and whether LD
allocation composes with a MIG instance, therefore remain unverified.
The $1.43\times$ figure is a substitute bound, not a measured MIG
speedup. Our experiment challenges its zero-interference assumption
but does not establish MIG's actual performance.

\item \textbf{Cold working sets show limited gains.}
After flushing L2, the compute-side gain for the measured GEMM
falls to approximately $1$--$5\%$ for FP8, while FP4 ranges from
a small slowdown to a similarly modest gain. Workloads with limited
data reuse should not be expected to reproduce the gains measured
with hot, resident data.

\item \textbf{The attention study has a limited scope.}
Our experiments use a custom decode-attention kernel on a single B300
platform with synthetic K/V data, rather than an end-to-end serving
system. Although we independently tune tile sizes across all 30
workload shapes, the evaluation does not cover the broader design
space of production attention implementations. Only four workloads
retain more than $5\%$ LD speedup after tuning. These findings
highlight the importance of kernel tuning, but do not rule out
potential benefits from LD placement in other attention
implementations or real-world serving workloads. SM clocks were
measured at 2032\,MHz at both session endpoints, with within-point
drift at or below 1.14\%.

\item \textbf{The die asymmetry remains unexplained.}
Primitive measurements show no substantial intrinsic difference
between the dies, but some collective configurations exhibit a
persistent performance asymmetry. We did not instrument the
rank-to-die mapping and kernel-level behavior needed to identify
its cause. A mapping-related explanation remains a hypothesis.

\item \textbf{Some comparisons use earlier measurements.}
The MLOPart and one-sided-copy results were collected in earlier
studies rather than rerun under the current experimental setup.
They provide qualitative comparisons, but differences in platform
and measurement conditions limit the precision of direct numerical
comparisons.

\end{enumerate}

Further work should prioritize kernel-level measurements of the
collective path. These could distinguish among L2-slice contention,
scheduling effects, fabric traffic, and other possible explanations
for the self-copy result. They could also establish whether the
residual die1 performance deficit comes from rank-to-die mapping
or another source. More broadly, evaluating LD on additional
platforms and resident-data workloads would test how widely the
measured latency--bandwidth trade-off and the interaction between
memory and compute placement generalize.


\bibliographystyle{plain}
\bibliography{ref}

\end{document}